\documentclass[8pt,prd,superscriptaddress,nofootinbib,notitlepage]{revtex4-1}

\usepackage{slashed}
\usepackage{caption}
\usepackage{subfigure}
\usepackage{amsmath,amssymb,graphicx,xcolor,mathtools}
\usepackage[colorlinks=true,linktocpage=green,linkcolor=blue,citecolor=blue]{hyperref}
\numberwithin{equation}{section}
\allowdisplaybreaks
\usepackage{hyperref}
\hypersetup{
	colorlinks=true,
	citecolor = blue,
	linkcolor= blue,
	filecolor= blue,      
	urlcolor= blue,
	pdftitle={HDL}
}
\usepackage{graphics}
\usepackage{mathrsfs}
\usepackage{simpler-wick}
\usepackage{dsfont}
\usepackage{mathtools}

\DeclareMathOperator{\Tr}{Tr}

\def\be {\begin{equation}}
	\def\ee {\end{equation}}
\def\bea {\begin{eqnarray}}
	\def\eea {\end{eqnarray}}
\def\bc {\begin{center}}
	\def\ec {\end{center}}
\def\nn {\nonumber}

\DeclareMathAlphabet{\mathpzc}{OT1}{pzc}{m}{it}

\def\sumintof{\sum\!\!\!\!\!\!\!\!\!\int\limits_{\{P\}}}

\def\sumintoff{\sum\!\!\!\!\!\!\!\!\!\int\limits_{\{p_{0}\}}}

\def\sumintob{\sum\!\!\!\!\!\!\!\!\!\int\limits_{P}}
\def\sumintof{\sum\!\!\!\!\!\!\!\!\!\int\limits_{\{P\}}}
\def\sumintofk{\sum\!\!\!\!\!\!\!\!\!\int\limits_{\{K\}}}

\def\nn{\nonumber\\}

\begin{document}
	
	\title{Thermodynamics of strongly magnetized dense quark matter from hard dense loop perturbation theory}

	\author{Sarthak Satapathy}
	\email{sarthaks680@gmail.com}
	\affiliation{ School of Physical Sciences, National Institute of Science Education and Research,
		An OCC of Homi Bhabha National Institute, Jatni-752050, India
	}
	
	\author{Sumit}
	\email{sumit@ph.iitr.ac.in}
\affiliation{School of Physics, Beijing Institute of Technology, Beijing 102488, China}
\affiliation{Department of Physics, Indian Institute of Technology Roorkee, Roorkee - 247667, India}	
	
	\author{Salman Ahamad Khan}
	\email{salmankhan.dx786@gmail.com}
	\affiliation{Department of Physics, Integral University, Lucknow - 226026, India
	}

\begin{abstract}

We discuss the hard dense loop perturbation theory approach for studying the thermodynamics of strongly magnetized dense quark matter. The free energy of quarks and gluons have been calculated for one-loop quark and gluon self-energies, respectively. The longitudinal and transverse components of pressure, magnetization, second-order quark number susceptibility, and speed of sound have been computed, and their behavior with chemical potential and magnetic field has been analyzed. Our numerical results show that the longitudinal pressure increases with chemical potential and magnetic field, while for the transverse component, it is diminished. We also analyze the longitudinal component of the speed of sound at high chemical potentials, which approaches the speed of light in the asymptotic limit. The obtained results may be helpful in studying magnetized quark matter in the core of neutron stars and magnetars.
	
\end{abstract}

\maketitle

\section{Introduction}
\label{sec:intro}

Quantum chromodynamics (QCD) exhibits two remarkable features known as confinement and asymptotic freedom at low and high energies, respectively. As a result of the asymptotic freedom, the hadronic matter gets deconfined in the quark and gluon degrees of freedom in the presence of extreme temperature and/or density. Such a state of matter is supposed to be present in the early universe just after the few microseconds of the big bang and also in the core of the compact stars. In the laboratory, it is created in the ultrarelativistic heavy ion collision (URHIC) experiments at the Relativistic Heavy Ion Collider (RHIC) at BNL and Large Hadron Collider (LHC) at CERN. The study of this extreme phase of matter has been central to the activities of the heavy ion physics community in recent years due to its great phenomenological relevance. Among many of its properties, the equation of state (EoS) is fundamental because it acts as an input to model the hydrodynamic evolution of hot and dense QCD matter. EoS also describes the thermodynamic properties and the QCD phase diagram; hence, there have been numerous efforts to compute the thermodynamics functions of this extreme phase of matter from the first principle using the nonperturbative lattice QCD simulations~\cite{Bazavov:2017dus, Bazavov:2017dsy}. Another approach to computing the thermodynamic functions is hard thermal loop perturbation theory which employs resummed propagators and vertices in place of bare ones and gives infrared improved and gauge-independent results at high temperatures and moderate densities~\cite{Haque:2012my, Andersen:2009tc, Andersen:2010ct, Andersen:2010wu, Andersen:2011sf, Andersen:2011ug, Haque:2013sja, Haque:2014rua}. The very high-density region of the QCD phase diagram is not accessible in lattice simulations due to the infamous sign problem. However, there 
have been efforts to explore the finite temperature and low-density region~\cite{Borsanyi:2012cr, Guenther:2017hnx, Bazavov:2017dus, DElia:2016jqh}. \par
Recently, robust evidence of the quark matter has been reported in the cores of the neutron stars~\cite{Annala:2019puf}, which motivates the study of the thermodynamic functions of high-density regions of the QCD phase diagrams. Such high-density and zero-temperature matter is also called cold quark matter. The EoS of cold quark matter was first obtained within perturbative QCD a long ago by Freedman and McLerran~\cite{Freedman:1977gz, Freedman:1976ub}, and later by Baluni~\cite{Baluni:1977ms} and Toimela~\cite{Toimela:1984xy}. Since then, there have been many efforts to improve the result systematically~\cite{Fraga:2004gz, Kurkela:2009gj, Blaizot:2000fc, Fraga:2001id, Fraga:2013qra, Kurkela:2014vha, Fraga:2015xha, Ghisoiu:2016swa, Annala:2017llu, Gorda:2018gpy, Gorda:2021kme}. Due to the presence of the magnetic field during the cosmological phase transitions in the early Universe, in the core of the neutron stars and the noncentral heavy ions collisions, it becomes a necessary exercise to include the effects of the magnetic field on the EoS of the strongly interacting matter~\cite{Adhikari:2024bfa}. The EoS has been studied at finite magnetic field using lattice QCD approach for zero~\cite{Bonati:2013vba, Levkova:2013qda, Bali:2014kia} as well as for finite baryon density~\cite{Astrakhantsev:2024mat} case. Attempts have also been made to study the speed of sound in the dense nuclear matter in a magnetic field through the Walecka model \cite{Mondal:2023baz} and thermodynamic properties of magnetized quark matter viz. speed of sound and isothermal compressibility via Nambu-Jona-Lasinio model \cite{Mondal:2024eaw}. In addition to it, the second-order fluctuations
of the baryon number, electric charge, and strangeness, which are related to EoS, in the external magnetic field were studied in~\cite{Ding:2021cwv, Borsanyi:2023buy}. The equation of state has also been studied using the perturbative thermal  QCD in the presence of weak as well as strong magnetic field~\cite{Rath:2017fdv, Bandyopadhyay:2017cle, Karmakar:2019tdp, Fraga:2023cef, Fraga:2023lzn}. Previously estimated in Ref. \cite{Huang:2011dc}, the authors in~\cite{Karmakar:2019tdp} have calculated the anisotropic pressures in the presence of the magnetic field. They have computed the longitudinal and transverse components of the pressure using the hard thermal loop perturbation theory (HTLpt) in the presence of a magnetic field where the magnetic field is the largest scale. All these studies have been performed on the finite temperature and finite density cases. To the best of our knowledge, no such study on the EoS has been carried out in the magnetic field at zero temperature and high density using a first principle perturbative approach in a magnetic field. The strong magnetic field might have a significant impact on cold QCD EoS since the dynamics of quarks in such large magnetic fields get Landau quantized and lead to a dimensional reduction in the dynamics from (3+1) dimensions to (1+1) dimensions. Contrary to the quarks, gluons are not directly affected by the strong magnetic field, but magnetic field dependence enters via the quark loop correction in the gluon self-energy. In particular, a better understanding of the thermodynamics of cold and dense matter in the presence of very strong magnetic fields is needed for the mass-radius relation of the magnetars where EoS acts as an input parameter~\cite{Kaspi:2017fwg}. In these compact stars~\cite{Schaffner:Compact}, the magnitude of the magnetic fields can be of the order of $10^{15}$ Gauss at the surface and possibly much higher in the core (up to $10^{20}$ Gauss~\cite{Ferrer:2010wz})hence, the inclusion of the strong magnetic field in the EoS of cold quark matter is essential to describe the structural properties of magnetars under such extreme conditions. Experimental efforts {\em like} NICER X-ray determination of mass and radius from millisecond pulsars PSR J0030+0451~\cite{Riley:2019yda, Miller:2019cac} and PSR J0740+6620~\cite{Riley:2021pdl, Miller:2021qha}, along with the LIGO-VIRGO detection of gravitational waves from binary neutron star mergers~\cite{LIGOScientific:2018cki, LIGOScientific:2018hze}, put a constrain on the equation of state that describes quark stars and neutron stars~\cite{MUSES:2023hyz}. Apart from the astrophysical relevance, high-density matter calculations have been rigorously analyzed in Refs. \cite{Gorda:2022yex,Osterman:2023tnt}. One of the other work for related studies has been done recently by Podo and Santoni \cite{Podo:2023ute}, which discusses the path integral formulation of dense fermions in magnetic fields. 
\par
In the present work, we have studied the thermodynamics of strongly magnetized cold quark matter using hard dense loop perturbation theory (HDLpt) up to one-loop. HDLpt is a reorganization of the conventional perturbative technique at very high chemical potential ($\mu$) and zero temperature~\cite{Manuel:1995td}. The naive one-loop computations are incomplete, as one-loop diagrams with soft $(\sim g\mu)$ external momenta and quarks in the internal lines with hard momenta $\mu$ are of the order of tree amplitudes. Therefore, we need to resum those diagrams. These diagrams are called hard dense loops (HDL) in analogy to the HTLpt. HDLpt has previously been used by Andersen {\it et al.} \cite{Andersen:2002jz} to study the thermodynamics of dense quark matter and subsequently use them to solve the Tolman-Oppenheimer-Volkov equations for obtaining the mass-radius relation for dense quark stars. In more recent work, this method was used by Fujimoto {\it et al.} \cite{Fujimoto:2020tjc} to calculate the pressure and speed of sound by including a nonzero mass in the computations. In this work, we have computed both quark and gluon contributions to the longitudinal and transverse components of the pressure using HDLpt in the strong magnetic field. Comparing our results with Ref. \cite{Fraga:2023lzn}, we see that the order of magnitude and behavior of the longitudinal pressure is the same as that obtained in our work. We have also computed the magnetization of the cold quark matter, which is required for the transverse pressure. Along the same lines, we have estimated the longitudinal and transverse contribution of second-order quark number susceptibility (QNS), which accounts for the fluctuations of quark number density. Further, we have also computed the speed of sound in the dense, strongly magnetized medium.  \par

The paper is organized as follows. In Sec. \ref{SE-Q}, we have derived the quark self-energy at finite density and vanishing temperature. In Sec. \ref{tp}, we have calculated the thermodynamic quantities viz. free energy of quarks from quark self-energy in Sec. \ref{FE}, the free energy of gluons through gluon self-energy in Sec. \ref{FG} respectively and the second-order quark number susceptibility along with the speed of sound has been discussed in Sec. \ref{cs}. The results for the said observables have been presented in Sec. \ref{Res}. Concluding discussions and outlook have been discussed in Sec. \ref{conc}. The dense sum integrals used to compute free energy have been provided in Appendix~\ref{App-A}.     

\section{Quark self-energy at high chemical potential in a strongly magnetized medium}
\label{SE-Q}
In this section, we will calculate the quark self-energy in a strongly magnetized medium at finite chemical potential using the HDL approximation. In the strong magnetic field at high temperature, the hierarchy of scales is given by $gT < T < \sqrt{q_f B}$, where $g$ is the coupling, $T$ is the temperature, $q_f$ is the charge of the fermion and $B$ is the magnetic field. In the presence of an external magnetic field, the energy of fermions gets quantized as $E_l = \sqrt{p_3^2 + m^2 + 2l|q_f B|}$, where $l \in [0, \infty)$. When the magnetic field is very strong, all the fermions are confined in the lowest Landau level (LLL), i.e., $l = 0$. This reduces the dynamics of the system from $(3+1)d$ to $(1+1)d$, where $d$ is the number of spacetime dimensions. At high $\mu$ and vanishing temperature i.e., $\mu / T \to \infty$, in the presence of magnetic field, the LLL condition is incorporated through the condition $l_{\text{max}} = \lfloor \mu^2/2|q_fB||\rfloor_{\mu^2 < 2|q_fB|} \to 0$, where $l_\text{max}$ is the maximum number of Landau levels and $\lfloor x \rfloor$ is the floor function. At vanishing temperatures when the system becomes degenerate, we apply HDLpt, where the external momentum is taken to be soft, i.e., $\sim g\mu$, and the loop momentum to be hard, i.e., $\sim \mu$. Thus, the fermions in the loop in the HDL approximation are of the order of the Fermi energy. Here, we will proceed by first computing the form factors of quark self-energy at finite $T$ and $\mu$ as done in Ref. \cite{Karmakar:2020mnj} and then taking the $T \to 0$ limit to include the HDL contributions. To start with, the general structure of quark self-energy in the massless limit in a strong magnetic field \cite{Karmakar:2019tdp} is given by 
\bea
\Sigma(p_0, p_3) = a(p_0, p_3) \slashed{u} +  b(p_0, p_3) \slashed{n} + c(p_0, p_3) \gamma_5\slashed{u} + d(p_0, p_3)\gamma_5\slashed{n}, 
\label{seq-1} 
\eea  
\begin{figure}[h]
	\centering
	\includegraphics[scale = 0.6]{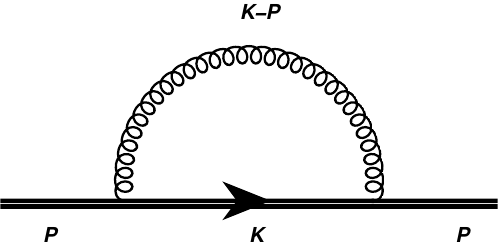}
	\caption{One-loop quark self-energy diagram in a strong magnetic field. The fermion propagator is shown by double straight lines, indicating a modification in a strong magnetic field.}
	\label{fig-FSE}
\end{figure}
where the notation $\slashed{V} = \gamma^\mu V_\mu$ for any 4-vector $V^\mu$, $u^\mu = (1,0,0,0)$ is the velocity of the heat bath, $n^\mu = (0,0,0,1)$ is the direction of the magnetic field which in our case is in the $z$-direction and $a,b,c,d$ are the form factors of the quark self-energy given by
\bea
&& a = \frac{1}{4}\Tr[\Sigma \slashed{u}],~~b = \frac{1}{4}\Tr[\Sigma \slashed{n}],~~ c = \frac{1}{4}\Tr[\gamma_5\Sigma \slashed{u}],~~ d = \frac{1}{4}\Tr[\gamma_5\Sigma \slashed{n}] .
\eea  
The one-loop quark self-energy has been shown in Fig. \ref{fig-FSE}. In the presence of a background magnetic field, the translational symmetry is broken, which is reflected in the coordinate space fermion propagator through the Schwinger phase factor \cite{Miransky:2015ava, Ghosh:2024hbf, Hattori:2023egw}. Had it been a closed fermion loop, gauge -or translational invariance would have been preserved as the sum of all Schwinger phases would have canceled out, as shown by Hattori {\it et al.} \cite{Hattori:2023egw}. For fermion self-energy, the breaking of translational symmetry is reflected in the coordinate space representation of the quark self-energy in the magnetic field as 
\bea
\Sigma(x,y) = -ig^2 \gamma^\mu S(x,y) \gamma_\mu \Delta(x - y),
\eea 
where $g$ is the coupling, $S(x,y)$ is the coordinate space quark propagator, which is not translationally invariant, and $\Delta(x-y)$ is the scalar part of the gluon propagator. We can write $S(x,y)$, that comes with an overall Schwinger phase factor $\exp(i\Phi(x,y))$, as 
\bea
S(x,y) = e^{i\Phi(x,y)} S(x-y), 
\eea  
where $$\Phi(x,y) = q_f \int_y^x d\xi^\mu \bigg[ A_\mu + \frac{1}{2}F_{\mu\nu}(\xi - y)^\nu \bigg], $$
Moreover, $S(x-y)$ is the translationally invariant part of the quark propagator. Physical quantities of interest should remain unaffected by the Schwinger phase factor, which implies that one should be able to gauge away this factor in the calculations. For this we note that the integrand in $\Phi(x,y)$, i.e. $ A_\mu + \frac{1}{2}F_{\mu\nu}(\xi - y)^\nu \equiv f_\mu(\xi;y)$, is curl-free or irrotational, as $\partial_\xi^\mu f^\nu(\xi; y) - \partial_\xi^\nu f^\mu(\xi; y) = 0,$
which implies that $\Phi(x,y)$ is independent of the path. By considering a straight-line path parameterized as 
\bea
\xi^\mu = y^\mu + \sigma(x - y)^\mu ,
\eea 
where $\sigma \in [0,1]$ and using the fact that $F^{\mu\nu}$ is antisymmetric, we get 
\bea
\Phi(x,y) = q_f \int_0^1 d\sigma A_\mu (x - y)^\mu .  
\eea 
For magnetic field in $z$-direction, if we consider the gauge potential in the symmetric gauge, i.e., $A_\mu(x) = \frac{B}{2}(0,-y,x,0)$, then we can perform a gauge transformation
\bea
A_\mu(\xi) \to A_\mu'(\xi) = A_\mu + \frac{\partial }{\partial\xi^\mu} \Lambda(\xi), 
\eea  
where 
\bea
\Lambda(\xi) = \frac{B}{2} (y'\xi_1 - x'\xi_2).
\eea 
This leads to 
\bea
\frac{\partial}{\partial \xi^\mu}\Lambda(\xi) = \frac{B}{2}(0, y', -x', 0), 
\eea 
following which $\Phi(x,y)$ vanishes and translational symmetry is recovered. The above calculations can be done for other gauge choices as well. On recovering translational invariance, we can write the fermion propagator as a Fourier transform, which is given by 
\bea
S_F(x-y) = \int \frac{d^4k}{(2\pi)^4} e^{-ik\cdot(x-y)}S(k),
\eea 
where $S(k)$ is the momentum space propagator given by 
\bea
\hspace{-0.75cm}iS(k) = \int_0^\infty ds \exp\bigg[ is(k_\parallel^2 - m_f^2 - \frac{k_\perp^2}{q_fBs} \tan(q_fBs))  \bigg] \bigg[(\slashed{k}_\parallel + m)(1 + \gamma^1\gamma^2 \tan(q_fBs)) - \slashed{k}_\perp (1 + \tan^2(q_f Bs)) \bigg].
\eea
The fermion propagator above can be expressed as a sum over the Landau levels for general strengths of the magnetic field as 
\bea
iS(k) =  ie^{-\frac{k_\perp^2}{q_fB}}  \sum_{l = 0}^\infty  \frac{ (-1)^l D_l(q_fB, k) }{k_\parallel^2 - m_f^2 + i\epsilon}
\eea 
where $l \in [0, \infty)$. Here $D_l(q_fB, k)$ is given by 
\bea
D_l(q_fB, k) = (\slashed{k}_\parallel + m)\bigg[ (1 - i\gamma^1 \gamma^2) L_l\bigg(\frac{2k_\perp^2}{q_fB}\bigg) - (1 + i\gamma^1 \gamma^2 ) L_{l-1}\bigg(\frac{2k_\perp^2}{q_fB}\bigg)  \bigg] - 4 \slashed{k}_\perp L_{l-1}^1\bigg(\frac{2k_\perp^2}{q_fB}\bigg),
\eea 
where $L_l^\alpha(x)$ is the generalized Laguerre polynomial given by 
\bea
(1 - z)^{-(1+ \alpha)} \exp\bigg(\frac{xz}{z - 1}\bigg) = \sum_{l = 0}^\infty  L_l^\alpha(x) z^l. 
\eea 
In a strong magnetic field, all the fermions occupy the LLL, due to which $l = 0$. In LLL, the Laguerre polynomials take the following values 
\bea
L_{-1}^1\bigg( \frac{2k_\perp^2}{q_fB} \bigg) = 0,~~ L_{-1}^0\bigg( \frac{2k_\perp^2}{q_fB} \bigg) = 0 \text{~and~} L_0\bigg( \frac{2k_\perp^2}{q_fB} \bigg) = 1.
\eea 
Therefore, the fermion propagator in LLL approximation is given by 
\bea
iS_F(k) = i e^{-\frac{k_\perp^2}{q_fB}}\frac{\slashed{k}_\parallel + m}{k_\parallel^2 - m^2} (1 - i\gamma_1\gamma_2) 
\eea 
and the quark self-energy can be written as
\bea
\Sigma(P) = -i g^2 C_F\int\frac{d^4k}{(2\pi)^4} \gamma_\mu S_F(K) \gamma^\mu \Delta(K-P),
\label{seq-3}
\eea 
where $C_F = (N_c^2 - 1)/(2N_c)$, $N_c$ is the number of colors, $\Delta(K-P)$ is the unmodified gluon propagator given by 
\bea
\hspace{-1cm}\Delta(K-P) = \frac{1}{ (k_0 - p_0)^2 - (k - p)^2  } = \frac{1}{ (K-P)_\parallel^2 - (k-p)_\perp^2  }.
\label{seq-4}
\eea  
In Eq. (\ref{seq-4}), the four-momentum has been divided into two parts given by $K^\mu = K^\mu_\parallel + K^\mu_\perp$, where $K^\mu_\parallel = (k_0, 0, 0, k_3)$ and $K^\mu_\perp = (0, k_1, k_2, 0)$. Using the expressions of $\Delta(K-P)$ and $S_F(K)$, the quark self-energy in Eq. (\ref{seq-3}) becomes
\bea
\Sigma(P) 
&=& -ig^2C_F \sum_f\int \frac{d^4K}{(2\pi)^4} e^{-k_\perp^2 / q_fB} \gamma_\mu \slashed{K}_\parallel  (\mathds{1} - i\gamma_1\gamma_2) \widetilde{\Delta}_\parallel(K) \Delta(K-P), 
\label{seq-6}
\eea 
where the bare quark masses have been neglected, and the index $f$ in the summation stands for the quark flavor. Also, $\widetilde{\Delta}_\parallel(K) = (1)/(k_0^2 - k_3^2)$. In evaluating Eq. (\ref{seq-6}), we have used the relation for the loop integration given by 
\bea
\int\frac{d^4K}{(2\pi)^4} \to iT \sum_{\{k_0\}}\int\frac{d^3k}{(2\pi)^3} \to iT \sum_{\{k_0\}}\int\frac{dk_3}{2\pi}\int\frac{d^2k_\perp}{(2\pi)^2}.
\eea 
Using Eq. (\ref{seq-6}) the form factors are given by 
\bea
a(p_0, p_3) &=& \frac{1}{4}\Tr[\Sigma \slashed{u}]  =  -\frac{g^2 C_F}{2}\sumintofk e^{-k_\perp^2/q_fB}\Tr\big[ \big(\mathds{1} + i\gamma_1\gamma_2\big) \slashed{K}_\parallel \slashed{u} \big]  \widetilde{\Delta}_\parallel(K)\Delta(K-P) \nn 
&=&  -2g^2C_F \sumintofk e^{-k_\perp^2/q_fB} k_0 \widetilde{\Delta}_\parallel(K)\Delta(K-P),
\label{seq-7a}
\eea 
\bea
b(p_0, p_3) &=& \frac{1}{4}\Tr[\Sigma \slashed{n}] = \frac{g^2 C_F}{2}\sumintofk e^{-k_\perp^2/q_fB}\Tr\big[ \big(\mathds{1} + i\gamma_1\gamma_2\big) \slashed{K}_\parallel \slashed{n} \big]  \widetilde{\Delta}_\parallel(K)\Delta(K-P) \nn 
&=&  2g^2C_F \sumintofk e^{-k_\perp^2/q_fB} k_3 \widetilde{\Delta}_\parallel(K)\Delta(K-P),
\label{seq-7b}
\eea 
\bea
c(p_0, p_3) &=& \frac{1}{4}\Tr[\gamma_5\Sigma \slashed{u}] =  -\frac{g^2 C_F}{2}\sumintofk e^{-k_\perp^2/q_fB}\Tr\big[ \gamma_5 \big(\mathds{1} + i\gamma_1\gamma_2\big) \slashed{K}_\parallel \slashed{u} \big]  \widetilde{\Delta}_\parallel(K)\Delta(K-P) \nn 
&=&  -2g^2C_F \sumintofk e^{-k_\perp^2/q_fB} k_3 \widetilde{\Delta}_\parallel(K)\Delta(K-P),
\label{seq-7c}
\eea
\bea
d(p_0, p_3) &=& \frac{1}{4}\Tr[\gamma_5\Sigma \slashed{n}] =  \frac{g^2 C_F}{2}\sumintofk e^{-k_\perp^2/q_fB}\Tr\big[ \gamma_5 \big(\mathds{1} + i\gamma_1\gamma_2\big) \slashed{K}_\parallel \slashed{n} \big]  \widetilde{\Delta}_\parallel(K)\Delta(K-P) \nn 
&=&  2g^2C_F \sumintofk e^{-k_\perp^2/q_fB} k_0 \widetilde{\Delta}_\parallel(K)\Delta(K-P). 
\label{seq-7d}
\eea 
From Eqs. (\ref{seq-7a})-(\ref{seq-7d}), we find that $a = -d$ and $b = -c$. To evaluate further, we employ HTL and perform the frequency sums to obtain the $T$ and $\mu$ dependence of the form factors. In the high dense limit we take the $T \to 0$ limit of $a(p_0, p_3)$ and $b(p_0, p_3)$. 
Further, for simplifying the calculations, we use the approximation $|K-P|_\perp < |K - P|_\parallel$, which would allow a geometric series expansion. Thus we can write $\Delta(K-P)$ as  
\bea
\Delta(K-P) &=& \frac{1}{ \big(K -P\big)_\parallel^2 - \big(k - p\big)_\perp^2  }  
= \frac{1}{\big( K - P \big)_\parallel^2}\bigg[ 1 - \frac{(k - p)_\perp^2}{ (K - P)_\parallel^2 }   \bigg]^{-1}  \nn 
&\approx& \Delta_\parallel(K - P)  +   (k - p)_\perp^2 \Delta_\parallel^2(K-P)  . 
\label{seq-7e}
\eea 
Following the above approximation $a(p_0, p_3)$ can be written as
\bea
a(p_0, p_3) &=& \frac{1}{4}\Tr[\Sigma \slashed{u}] 
= -2g^2 C_F \int\frac{d^3k}{(2\pi)^3} e^{-k_\perp^2/ q_fB}\big[ \mathcal{N}_1 + (k-p)_\perp^2 \mathcal{N}_2 \big] , 
\label{seq-8}
\eea  
where $\mathcal{N}_1$ and $\mathcal{N}_2$ are given by $$\mathcal{N}_1 = T\sum\frac{k_0}{(K_\parallel^2)(K-P)_\parallel^2},~~~\mathcal{N}_2 = T\sum\frac{k_0}{K_\parallel^2(K-P)_\parallel^4} = -\frac{1}{2k_3}\frac{\partial \mathcal{N}_1}{\partial p_3}.$$ 
On performing the transverse momentum integration in Eq. (\ref{seq-8}), $a(p_0, p_3)$ comes out to be
\bea
a = -g^2 C_F \frac{|q_fB|}{4\pi^2} \int_{-\infty}^{+\infty} dk_3 \big[\mathcal{N}_1 + \big(p_\perp^2 + q_fB\big)\mathcal{N}_2  \big] .
\label{a}
\eea  
After performing the Matsubara frequency sums \cite{Laine:2016hma,Haque:2024gva}, $\mathcal{N}_1$ and $\mathcal{N}_2$ are given by
\bea
&& \mathcal{N}_1 = -\frac{1}{4k_3}\bigg[\frac{n_B(k_3) + n_F(k_3 - \mu)}{p_0 + p_3} + \frac{n_B(k_3) + n_F(k_3 + \mu)}{p_0 - p_3} \bigg],~~ \mathcal{N}_2 = -\frac{1}{2k_3}\frac{\partial \mathcal{N}_1}{\partial p_3} . 
\eea 
In the dense limit, i.e., $T \to 0$, the Bose-Einstein distribution function $n_B(k_3)$ and the antiparticle distribution function $n_F(k_3 + \mu)$ do not contribute. Further, the Fermi-Dirac distribution function $n_F(k_3 - \mu)$, becomes a Heaviside step function i.e., $ \displaystyle{\lim_{T \to 0}} n_F(k_3 - \mu) \to \Theta(\mu - |k_3|)$, where $\mu$ is the Fermi momentum. The form factor $a(p_0, p_3)$ consists of two integrations as given in Eq. (\ref{a}). The first integration is calculated as follows 
\bea
\hspace{-1cm}\displaystyle{\lim_{T \to 0}} \int_{-\infty}^{+\infty } dk_3 \mathcal{N}_1 &=& -\displaystyle{\lim_{T \to 0}}  \bigg(\frac{e^{\gamma_E} \Lambda^2}{4\pi}\bigg)^\epsilon\int_{-\infty}^{+\infty}\frac{dk_3}{4k_3}  k_3^{-2\epsilon}  \frac{n_F(k_3-\mu)}{p_0 + p_3}  
=  - \bigg(\frac{e^{\gamma_E} \Lambda^2}{4\pi}\bigg)^\epsilon \int_{0}^{\mu}\frac{dk_3}{4k_3}k_3^{-2\epsilon}\frac{\Theta(\mu - |k_3|)}{p_0 + p_3},
\label{N_1}
\eea
where $2\epsilon =  1 - d$, $d$ being the number of spatial dimensions, $\big(\frac{e^{\gamma_E} \Lambda^2}{4\pi}\big)^\epsilon$ is the regularization factor introduced to regulate the UV divergences, and $\Lambda$ is an arbitrary momentum scale. By substituting the above expression in Eq. (\ref{N_1}) and carrying out the integration we get 
\bea 
\displaystyle{\lim_{T \to 0}} \int_{-\infty}^{+\infty } dk_3 \mathcal{N}_1 &=& \frac{1}{8(p_0 + p_3)\epsilon} +  \frac{(p_0 - p_3)}{8P_\parallel^2} \Big[\gamma_E -  \log 4\pi - 2 \log\left|\frac{\mu}{\Lambda}\right| \Big] + \mathcal{O}(\epsilon),
\label{N_11}
\eea 
where $P_\parallel^2 = p_0^2 - p_3^2$. The other integration involving $\mathcal{N}_2$ is given by
\bea
\displaystyle{\lim_{T \to 0}} \int_{-\infty}^{+\infty } dk_3~|q_fB|~\mathcal{N}_2 = \frac{|q_fB|}{8\mu}\frac{(p_0 - p_3)^2}{P_\parallel^4} + \mathcal{O}(\epsilon).
\label{N_22}
\eea 
Thus, $a(p_0, p_3)$, obtained by combining Eqs. (\ref{N_11}) and (\ref{N_22}) and substituting them in Eq. (\ref{a}) is given by 
\bea
a(p_0, p_3) = -g^2C_F\frac{|q_fB|}{4\pi^2} \bigg[ \frac{1}{8(p_0 + p_3)\epsilon} +  \frac{(p_0 - p_3)}{8P_\parallel^2} \bigg[\gamma_E -  \log 4\pi - 2 \log\left|\frac{\mu}{\Lambda}\right|~\bigg]  + \frac{|q_fB|}{8\mu}\frac{(p_0 - p_3)^2}{P_\parallel^4}  \bigg] + \mathcal{O}(\epsilon). 
\label{a-final}
\eea 
The form factor $b(p_0, p_3)$ is given by 
\bea
b(p_0, p_3) = g^2C_F\frac{|q_fB|}{4\pi^2}\int_{-\infty}^{+\infty} dk_3~k_3\big[\mathcal{N}_3 + |q_fB|\mathcal{N}_4\big] ,
\label{b}
\eea 
where $$\mathcal{N}_3 = \frac{1}{4k_3^2}\bigg[ \frac{n_B(k_3) + n_F(k_3-\mu)}{p_0 + p_3} - \bigg\{\frac{n_B(k_3) + n_F(k_3 +\mu)}{p_0 - p_3}\bigg\}  \bigg],~~~\mathcal{N}_4 = -\frac{1}{2k_3}\frac{\partial  \mathcal{N}_3}{\partial p_3}.$$ 
The integrations can be carried out by following the calculations in $a(p_0, p_3)$. They are given by  
\bea
\displaystyle{\lim_{T \to 0}}\int_{-\infty}^{+\infty}dk_3~k_3~\mathcal{N}_3  
&=& \displaystyle{\lim_{T \to 0}}\bigg(\frac{e^{\gamma_E} \Lambda^2}{4\pi}\bigg)^\epsilon  \int_{-\infty}^{+\infty}\frac{dk_3}{4k_3} k_3^{-2\epsilon}\frac{n_F(k_3-\mu)}{p_0 + p_3}  
= \bigg(\frac{e^{\gamma_E}\Lambda^2}{4\pi}\bigg)^\epsilon  \int_0^\mu\frac{dk_3}{4k_3}k_3^{-2\epsilon}\frac{\Theta(\mu - |k_3|)}{p_0 + p_3}   \nn 
&=& -\frac{1}{8(p_0 + p_3)\epsilon} - \frac{(p_0 - p_3)}{8P_\parallel^2}\bigg[\gamma_E - \log 4\pi - 2\log\left|\frac{\mu}{\Lambda}\right| \bigg] + \mathcal{O}(\epsilon)
\label{N_3}
\eea  
 and 
\bea
\hspace{-0.5cm}\displaystyle{\lim_{T \to 0}}\int_{-\infty}^{+\infty} dk_3~k_3~|q_fB|~\mathcal{N}_4 
= -\frac{|q_fB|}{8} \frac{\partial}{\partial p_3} \bigg(\frac{e^{\gamma_E}\Lambda^2}{4\pi} \bigg)^\epsilon  \int_0^\mu \frac{dk_3}{k_3^2} k_3^{-2\epsilon} \frac{\Theta(\mu - |k_3|)}{p_0 + p_3} = -\frac{|q_fB|}{8\mu}\frac{(p_0 - p_3)^2}{P_\parallel^4} + \mathcal{O}(\epsilon) . 
\label{N_4}
\eea 
By combining Eqs. (\ref{N_3}) and (\ref{N_4}) and substituting them in Eq. (\ref{b}) we obtain $b(p_0, p_3)$ as
\bea
b(p_0, p_3) = g^2 C_F\frac{|q_fB|}{4\pi^2}\bigg[ \frac{1}{8(p_0 + p_3)\epsilon} + \frac{(p_0 - p_3)}{8P_\parallel^2}\bigg(\gamma_E - \log 4\pi - 2\log\left|\frac{\mu}{\Lambda}\right|\bigg) + \frac{|q_fB|}{8\mu}\frac{(p_0 - p_3)^2}{P_\parallel^4}   \bigg] + \mathcal{O}(\epsilon).
\label{b-final}
\eea 
In the above expressions of $a(p_0, p_3)$ and $b(p_0, p_3)$, we see that there are divergent terms present, which are of the $\mathcal{O}(\epsilon^{-1})$. These vacuum contributions can be absorbed in the zero-temperature renormalization, and the final expression of the form factors is given by
\bea
a(p_0, p_3) = -b(p_0, p_3) = -g^2C_F\frac{|q_fB|}{4\pi^2} \bigg[  \frac{(p_0 - p_3)}{8P_\parallel^2} \bigg(\gamma_E -  \log 4\pi - 2 \log\left|\frac{\mu}{\Lambda}\right|~\bigg)  + \frac{|q_fB|}{8\mu}\frac{(p_0 - p_3)^2}{P_\parallel^4}  \bigg] + \mathcal{O}(\epsilon),
\label{ab_final}
\eea 
where we see that $a(p_0, p_3)$ is the negative of $b(p_0, p_3)$.

\section{Thermodynamics}
\label{tp}

\subsection{Free energy of quarks}
\label{FE}
In this section, we present the calculation of the quark-free energy in a strong magnetic field using the form factors calculated in Sec. \ref{SE-Q}. The free energy of quarks has been calculated for the strongly magnetized medium in high-temperature scenario~\cite{Karmakar:2019tdp}. We briefly recapitulate those calculations in order to calculate the quark-free energy for dense strongly magnetized medium utilizing the form factors at zero temperature and finite density, which have been mentioned in Eq. (\ref{ab_final}). The free energy of the $f$ flavor quark can be written as 
\bea
F_{f} = -N_c\sumintof \ln\big(\det \big[ S_{\text{eff}}^{-1}(p_0, p_3) \big] \big).
\label{fe-1}
\eea 
where $S_\text{eff}^{-1}$ is the inverse of the effective fermion propagator. 
Here, the sum integral is given by 
\bea
\sumintof \equiv T \sum_{\{p_0\}} \int \frac{d^3p}{(2\pi)^3}.
\eea 
Also, the subscript $f$ stands for the flavor of the quark, which we considered $u$ and $d$ in this work, 
For calculating $S_\text{eff}^{-1}$, we will first focus on the effective propagator in coordinate space, which can be written as a matrix element of the form 
\bea
S_\text{eff}(u,u') = i \langle u|[ \gamma^\mu \Pi_\mu + \Sigma  ]^{-1}  |u' \rangle,  
\label{seff-1}
\eea     
where $u = (t,x,y,z)$ and  $\Pi^\mu $ is the canonical momentum defined as $\Pi^0 = i\partial_t + \mu$ and $\Pi^k = i\partial^k + gA^k$, where $k = x,y,z$ and $\Sigma$ is the fermion self-energy in (\ref{seq-1}). We consider the gauge potential in the Landau gauge, i.e., $A^\mu = (0, 0, Bx, 0)$, following which the modified self-energy is given by  
\bea
\widetilde{\Sigma} = -\gamma^2 \Sigma \gamma^2 = -\Sigma. 
\label{seff-11} 
\eea 
The matrix element in Eq. (\ref{seff-1}) can be rewritten as 
\bea
S_\text{eff}(u,u') 
&=& i \langle u| [ \slashed{\Pi} +  \widetilde{\Sigma}  ]  [ \slashed{\Pi}^2 -  \Sigma^2 ]^{-1}  |u' \rangle  . 
\eea   
The Fourier transform of $S_\text{eff}(u,u')$ along the translationally invariant directions, $t$ and $z$, is given by 
\bea
S_\text{eff}(p_0, p_3; u_\perp, u_\perp') = \int dt~dz~e^{ip_0(t-t') - ip_3(z - z') }S_\text{eff}(u, u'),
\label{seff-2}
\eea
where 
\bea S_\text{eff}(p_0, p_3; u_\perp, u_\perp') = e^{i\Phi(u_\perp, u'_\perp)} S_\text{eff}(p_0, p_3; u_\perp - u_\perp'),~~~u_\perp = (x,y).
\label{seff-3}
\eea
We see that the Schwinger phase factor $e^{i\Phi(u_\perp, u'_\perp)}$ breaks the translational invariance and $S_\text{eff}(p_0, p_3; u_\perp - u_\perp')$ is the translationally invariant part of the propagator. Here $\Pi^\mu$ can be broken down into $\Pi^0$, $\Pi^\perp$ and $\Pi^3$, and in the LLL ($l = 0$), the eigenvalue of $\Pi^\perp = \sqrt{2l|q_fB|}|_{l = 0} = 0$. Thus, we can write the propagator as  
\bea
S_\text{eff}(p_0, p_3; u_\perp, u_\perp') = \langle u_\perp | [ \gamma^0 p_0 - \gamma^3 p_3 + \widetilde{\Sigma}  ] [p_0^2 - p_z^2 - \Sigma^2 ]^{-1} |  u'_\perp \rangle.
\eea   
The matrix element can be rewritten by using the spectral expansion of the unit operator 
\bea
\int dp \langle u_\perp|p \rangle \langle r | u_\perp' \rangle = \int_{-\infty}^{+\infty} dp~ \psi_p(u_\perp) \psi^*_p(u_\perp') = \delta^2(u_\perp - u_\perp'), 
\label{spec-exp}
\eea 
where $\psi_p(u_\perp), \psi^*_p(u_\perp')$ are the normalized wave functions.
Thus $S_\text{eff}(p_0, p_3; u_\perp, u_\perp')$ is given by 
\bea
S_\text{eff}(p_0, p_3; u_\perp, u_\perp')  &=&  \int_{-\infty}^{+\infty} dp~ \langle u_\perp | p \rangle   [ \gamma^0 p_0 - \gamma^3 p_3 + \widetilde{\Sigma}  ] [p_0^2 - p_3^2 - \Sigma^2 ]^{-1}  \langle p | u_\perp' \rangle  \nn 
&=&  \frac{q_fB}{2\pi}   \exp( i\Phi(u_\perp, u_\perp')  )  \exp\bigg( -q_fB\frac{(u_\perp - u_\perp')^2}{4}  \bigg)\bigg[ \frac{ (\gamma\cdot p)_\parallel + \widetilde{\Sigma} }{  (\gamma\cdot p)_\parallel^2 - \Sigma^2 }  \bigg ], 
\label{seff-4}  
\eea  
where 
\bea
\Phi(u_\perp, u'_\perp) = \frac{q_fB}{2}(x + x')(y - y').
\eea 
From Eq. (\ref{seff-3}), by considering the translationally invariant part of the propagator and performing Fourier transform of Eq. (\ref{seff-4}) with respect to the transverse spatial coordinates,
\bea
S_\text{eff}(p_0, p_z) &=& \int d^2u_\perp~e^{-ip_\perp\cdot u_\perp}  S_\text{eff}(p_0, p_z; u_\perp ) \nn 
&=& \frac{q_fB}{2\pi}\int d^2u_\perp~\exp\bigg(-p_\perp\cdot u_\perp -q_fB\frac{u_\perp ^2}{4}  \bigg) \bigg[ \frac{ (\gamma\cdot p)_\parallel + \widetilde{\Sigma} }{  (\gamma\cdot p)_\parallel^2 - \Sigma^2 }  \bigg ] 
= 2e^{-\frac{p_\perp^2 }{ q_fB}}  \bigg[ \frac{ (\gamma\cdot p)_\parallel + \widetilde{\Sigma} }{  (\gamma\cdot p)_\parallel^2 - \Sigma^2 }  \bigg ]. 
\label{seff-5}
\eea   
Following this, we can also write the inverse of the effective fermion propagator $S_\text{eff}^{-1}(p_0, p_3)$ in coordinate space as a matrix element, using Eq. (\ref{seff-1}) and (\ref{seff-11}) as
\bea
S_\text{eff}^{-1}(u,u') = -i\langle u| \gamma^\mu \Pi_\mu + \Sigma   |u' \rangle,
\eea 
which can be written in momentum space as
\bea
S_\text{eff}^{-1}(p_0, p_3) = \gamma^0p_0 - \gamma^3p_3 + \Sigma(p_0,p_3) = (p_0 + a)\gamma^0 + (b - p_3)\gamma^3 + c\gamma_5\gamma^0 + d\gamma_5\gamma^3.   
\eea 
The determinant of the $S_{\text{eff}}^{-1}(p_0, p_3)$ comes out to be
\bea
\det \big[ S_{\text{eff}}^{-1}(p_0, p_3) \big] &=& P_\parallel^4\bigg[ 1 + \frac{4a^2 - 4b^2 + 4ap_0 + 4bp_3}{P_\parallel^2} \bigg].
\label{fe-3}
\eea 
Utilizing the above expression of Eq.~\eqref{fe-3}, the free energy of quarks becomes
\bea
F_{f} &=& -2N_c\sumintof \ln P_\parallel^2 - N_c\sumintof \ln\bigg[ 1 + \frac{4a^2 - 4b^2 + 4ap_0 + 4bp_3}{P_\parallel^2}  \bigg] = F_{0,f} + F^{\prime}_{f} ,
\label{fe-4}
\eea 
where $F_{0,f}$ is the ideal free energy and $F_{f}^{\prime}$ is the one-loop correction to the free energy. An expansion of $F_{f}^{\prime}$ upto $\mathcal{O}(g^4)$ terms reads
\bea
F_{f}^{\prime} = -N_c\frac{|q_fB|}{(2\pi)^2}\sumintoff dp_3 \bigg[ \frac{4 (ap_0 + bp_3)}{P_\parallel^2} + \frac{4\big(a^2 P_\parallel^2 - b^2 P_\parallel^2 - 2a^2 p_0^2 - 2b^2 p_3^2 - 4ab p_0p_3\big)}{P_\parallel^4}  \bigg] + \mathcal{O}(g^6),
\label{fe-5}
\eea  
where the above expansion is valid for $g^2\big(q_fB/\mu^2\big) < 1$, where $g << 1$ and $q_fB/\mu^2 \gtrsim 1$, a condition valid for strong magnetic field. Substituting the expressions of $a(p_0, p_3)$ and $b(p_0, p_3)$ in $F_f'$ we get 
\bea
F_{f}^{\prime}(\mu, q_fB)   
&=& -N_c  \frac{|q_fB|}{(2\pi)^2}  \bigg[  -\frac{1}{8}\bigg(\frac{|q_fB| }{\mu}\bigg)^2 \bigg( g^2C_F\frac{|q_f B|}{4\pi^2} \bigg)^2\big\{ \mathcal{I}_{630} + \mathcal{I}_{603} - \mathcal{I}_{612} - \mathcal{I}_{621}  \big\} -\frac{1}{2} \bigg(g^2C_F\frac{|q_f B|}{4\pi^2} \bigg) \log\bigg(\frac{e^{\gamma_E}\Lambda^2}{4\pi\mu^2}\bigg) \nn
&& \times  \big\{  \mathcal{I}_{210} - \mathcal{I}_{201} \big\} -\frac{1}{8}  \bigg(g^2C_F\frac{|q_f B|}{4\pi^2} \bigg)^2 \bigg\{ \log\bigg(\frac{e^{\gamma_E}\Lambda^2}{4\pi\mu^2}\bigg)  \bigg\}^2 \big\{ \mathcal{I}_{420} - 2\mathcal{I}_{411} + \mathcal{I}_{402} \big\}\bigg], 
\label{I_abc}
\eea 
where $\mathcal{I}_{\alpha\beta\gamma} \equiv \mathcal{I}_{\alpha\beta\gamma}(\mu)$ are the dense sum integrals, which are basically the $T \to 0$ limit of the sum integrals at finite temperature and density. These sum integrals were first studied in a seminal paper by Gorda {\it et al.} \cite{Gorda:2022yex}, where it was shown that the limits $T = 0$ and $T \to 0$ are not equivalent for such dense sum integrals. Following Ref. \cite{Gorda:2022yex}, we have derived the general structure of $\mathcal{I}_{\alpha\beta\gamma}$ in Appendix \ref{App-A} given by 
\bea  
\hspace{-0.55cm}\mathcal{I}_{\alpha\beta\omega}(\mu) =  \displaystyle{\lim_{T \to 0}}\sumintof  \frac{ p_0^{2\beta} p^{2\omega} }{ P^{2\alpha} }  
= \bigg( \frac{e^{\gamma_E} \Lambda^2}{4\pi}  \bigg)^\epsilon \frac{i\mu}{2\pi}\frac{ \Gamma(\alpha - \omega - d/2) \Gamma(d/2 + \omega) (i\mu)^{d + 2\omega - 2\alpha + 2\beta} }{ (4\pi)^{d/2} \Gamma(\alpha) \Gamma(d/2) \big( 1 + d + 2\omega - 2\alpha + 2\beta  \big)  } \Big\{ \big( -1\big)^{d + 2\omega - 2\alpha + 2\beta} -1 \Big\}. 
\label{SI}
\eea 
In Eq. (\ref{SI}), $p_0 = ip_n = i(2n + 1)\pi T + \mu,$ $n\in Z$, is the fermionic Matsubara frequency and $p$ is the spatial momentum. Following Eq. (\ref{SI}), the dense sum integrals have been calculated, and their expressions have been provided in Appendix \ref{App-A}. The ideal free energy of the $f$ flavor quark is given by
\bea
F_{0,f}(\mu, q_fB) = -2N_c\sumintof\frac{d^3p}{(2\pi)^3}\ln P_{\parallel}^2  
=  -N_c\frac{|q_fB|}{4\pi^2}\mu^2 .
\label{F_Ideal}
\eea 
The quark-free energy carries a tree-level contribution \cite{Podo:2023ute} due to the magnetic field, i.e., $-B^2/2$, which does not contribute to any medium-dependent corrections. Adding this term, the expression of free energy in Eq. (\ref{fe-4}) can be rewritten as  
\bea
F_f(\mu, q_fB) = F_f^{\prime} + F_{0,f} - \frac{B^2}{2}.
\label{F_HDL}
\eea 
Thus, the longitudinal pressure $P_L$ and the transverse pressure $P_\perp$ are given by 
\bea
P_L = -F_f,~~P_\perp = - F_f - eB\cdot \mathcal{M},
\eea
where $\mathcal{M} = -\frac{\partial F_f}{\partial (eB)}$ is the magnetization per unit volume. 
\subsection{Free energy of gluons}
\label{FG}
In this section, we calculate the gluon free energy contribution by calculating the one-loop gluon self-energy in the magnetic field in the $T \to 0$ limit at finite $\mu$. The one-loop gluon self-energy consists of four diagrams, out of which, the diagram with fermion loop has a nonvanishing contribution at finite density and magnetic field, and the gluon and ghost loops do not contribute in the following limit. In order to calculate the free energy, we will proceed in a manner similar to that done for quarks, i.e., by computing the form factors of the gluon self-energy. A general structure of the gauge boson self-energy $\Pi^{\mu\nu}$ at finite temperature and magnetic field has been constructed in Ref. \cite{Karmakar:2018aig}, which we briefly review in this section. In vacuum, the gluon self-energy $\Pi^{\mu\nu}$ is given by 
$$\Pi^{\mu\nu}(P) = V^{\mu\nu}\Pi(P^2),$$  
where $V^{\mu\nu} = g^{\mu\nu} - \frac{P^\mu P^\nu}{P^2}$ is the vacuum projection vector, $\Pi(P^2)$ is a Lorentz scalar, $g^{\mu\nu} = (1,-1,-1,-1)$ and $P^\mu = (p_0, p_1, p_2, p_3)$. It satisfies the transversality condition $P_\mu \Pi^{\mu\nu} = 0$ and is symmetric in Lorentz indices i.e., $\Pi^{\mu\nu} = \Pi^{\nu\mu}$. At finite $T$ or $\mu$, Lorentz symmetry is broken due to the presence of the heat bath four-vector $u^\mu$, given by $u^{\mu} = (1,0,0,0)$ and thus $P^\mu$ can be decomposed into parallel and perpendicular components, relative to $u^\mu$. They are given by
$$P^\mu_\parallel = (P\cdot u)u^\mu,~~~P^\mu_\perp = \widetilde{P}^\mu = P^\mu - (P\cdot u)u^\mu~~~\text{and~~} \widetilde{g}^{\mu\nu} = g^{\mu\nu} - u^\mu u^\nu .$$ One can also consider the parallel and perpendicular direction components of $u^\mu$ with respect to the unit direction of $P^\mu$. They are given by
$$u^\mu_\parallel =  \frac{(u\cdot P)}{P_\mu P^\mu} P^\mu = \frac{(u\cdot P)}{P^2}P^\mu,~~~u^\mu_\perp = \overline{u}^\mu = u^\mu - u_\parallel^\mu = u^\mu - \frac{p_0 }{P^2}P^\mu.$$
Similarly to $V^{\mu\nu}$ one can construct two independent second rank projection tensors $A^{\mu\nu}$ and $B^{\mu\nu}$ which are orthogonal to each other given by
\bea
A^{\mu\nu} = \widetilde{g}^{\mu\nu} - \frac{ \widetilde{P}^\mu\widetilde{P}^\nu }{\widetilde{P}^2}~~\text{and}~~
B^{\mu\nu} = \frac{\overline{u}^\mu \overline{u}^\nu}{\overline{u}^2},
\label{FG1}
\eea 
and their combination gives $V^{\mu\nu}$. Following the above discussions, the covariant form of $\Pi^{\mu\nu}$ is given by 
\bea
\Pi^{\mu\nu} = \Pi_T A^{\mu\nu} + \Pi_L B^{\mu\nu} ,
\label{FG2} 
\eea 
where $\Pi_T$ and $\Pi_L$ are the transverse and longitudinal form factors of $\Pi^{\mu\nu}$. At finite temperatures and magnetic fields, both the Lorentz and rotational symmetry of the system are broken. A new four-vector $n^\mu$, along the direction of the magnetic field, comes into the picture, which is taken to be in the $z$ direction for simplicity purposes. Due to the breaking of rotational symmetry, the four-momentum is divided into two parts $P^\mu_\parallel = (p_0, 0, 0, p_3)$ and $P^\mu_\perp = (0, p_1, p_2, 0)$. Now there are the vectors and tensors associated with the system are $P^\mu, u^\mu$ and $n^\mu$ along with $g^{\mu\nu}$. Similar to the finite temperature case, the components of four-momentum and metric tensors that are parallel and perpendicular to $u^\mu$ and $n^\mu$ are given by 
\bea
P^\mu_\parallel = (P\cdot u)u^\mu + (P\cdot n)n^\mu,~~P^\mu_\perp = P^\mu - P^\mu_\parallel,~~g^{\mu\nu}_\parallel = u^\mu u^\nu - n^\mu n^\nu,~\text{and~} g^{\mu\nu}_\perp = g^{\mu\nu} - g^{\mu\nu}_\parallel.
\label{FG3}
\eea 
Along with $A^{\mu\nu}$ and $B^{\mu\nu}$, here we will have additional second-rank tensors. The projection of $A^{\mu\nu}$ along $n^\mu$ is given by $A^{\mu\nu}n_\nu$.
Similar to $B^{\mu\nu}$ in Eq. (\ref{FG1}), we can construct a tensor orthogonal to both $P^\mu$ and $B^{\mu\nu}$ as 
\bea
Q^{\mu\nu} = \frac{\overline{n}^\mu  \overline{n}^\nu}{\overline{n}^2}.
\label{FG4}
\eea 
A third projection tensor $R^{\mu\nu}$ can be constructed such that the sum of $R^{\mu\nu}, B^{\mu\nu}, Q^{\mu\nu}$ gives the vacuum projection operator $V^{\mu\nu}$. Thus $R^{\mu\nu}$ is given by
\bea
R^{\mu\nu} &=& V^{\mu\nu} - B^{\mu\nu} - Q^{\mu\nu} = A^{\mu\nu} - B^{\mu\nu} =  g^{\mu\nu}_{\perp} - \frac{P^\mu_\perp  P^\nu_\perp}{P_\perp^2} .
\eea 
The tensors $B^{\mu\nu}, R^{\mu\nu}, Q^{\mu\nu}$, if we collectively call them $\mathcal{K}^{\mu\nu}$, satisfy the relations
\bea
P_\mu \mathcal{K}^{\mu\nu} = 0,~~
\mathcal{K}^{\mu\lambda} \mathcal{K}^{\nu}_{~\lambda} = \mathcal{K}^{\mu\nu},~~ 
\mathcal{K}^{\mu\nu} \mathcal{K}_{\mu\nu} = 1,
\eea 
and are orthogonal to each other
\bea
\mathcal{K}^{\mu\nu} \mathcal{K}'_{\mu\nu} = 0,\text{~where~} \mathcal{K}\neq \mathcal{K}'
\eea 
The fourth tensor $N^{\mu\nu}$ is given by 
\bea
N^{\mu\nu} = \frac{ \overline{u}^\mu \overline{n}^\nu + \overline{u}^\nu \overline{n}^\mu  }{ \sqrt{\overline{u}^2} \overline{n}^2},
\eea 
which satisfies the relations 
\bea
N^{\mu\rho} N_{\rho\nu} = B^\mu_\nu + Q^\mu_\nu,~~~ B^{\mu\rho}N_{\rho\nu} + N^{\mu\rho}B_{\rho\nu} = N^\mu_\nu \nn
Q^{\mu\rho}N_{\rho\nu} + N^{\mu\rho}Q_{\rho\nu} = N^\mu_\nu,~~~ R^{\mu\rho}N_{\nu\rho} = N^{\mu\rho}R_{\nu\rho}  = 0  .
\eea 
The general covariant structure of $\Pi^{\mu\nu}$ can be written as 
\bea
\Pi^{\mu\nu} = bB^{\mu\nu}  + cR^{\mu\nu} +  dQ^{\mu\nu} + aN^{\mu\nu},
\eea 
where the form factors are 
\bea
 a = \frac{1}{2}N^{\mu\nu}\Pi_{\mu\nu},~~
 b = B^{\mu\nu}\Pi_{\mu\nu},~~ 
 c = R^{\mu\nu}\Pi_{\mu\nu},~~ 
 d = Q^{\mu\nu}\Pi_{\mu\nu} .
\eea 
In general, the form factors are dependent on temperature and magnetic field. The $\Pi^{\mu\nu}$ can also be written as 
\bea
\Pi^{\mu\nu} = \Pi^{\mu\nu}_g + \Pi^{\mu\nu}_s  ,
\eea 
\begin{figure}[h]
	\centering
	\includegraphics[scale = 0.6]{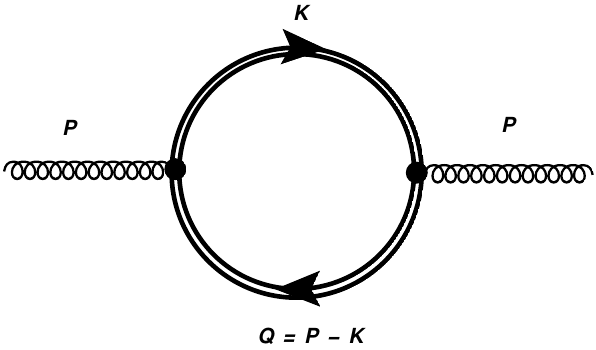}
	\caption{One-loop gluon self-energy diagram with fermion lines in a strong magnetic field. The fermion propagator is shown by double straight lines, indicating a modification in a strong magnetic field.}
	\label{fig-GSE}
\end{figure}
where $\Pi^{\mu\nu}_g$ is the contribution of ghost and gluon loops and $\Pi^{\mu\nu}_s$ is the contribution from the fermion loop diagram shown in Fig. \ref{fig-GSE}. At $T \to 0, \mu, B \neq 0$, $\Pi^{\mu\nu}_g$ has no contribution. Therefore, we will consider only the contribution of $\Pi^{\mu\nu}_s$. In the $T \to 0$ limit, the temperature dependent contributions in $\Pi^{\mu\nu}_s$ are neglected. Here we would like to discuss about the Debye mass $m_D$ in strong magnetic field at finite $T$ and $\mu$, given by 
\bea
m_D^2 = m_{D,g}^2 + m_{D,q}^2, 
\eea 
where $m_{D,g}$ and $m_{D,q}$ are the Debye mass coming from the gluonic loop and quark loop, respectively. The gluon contribution, unaffected by the magnetic field, is given by 
\bea
m_{D,g}^2(T) = N_c\frac{g^2 T^2}{3}.
\eea 
In the strong magnetic field limit, we have considered massless quarks. Their contribution to the Debye mass is given by
\bea
m_{D,q}^2(T, \mu, q_fB) &=& \sum_f \frac{g^2 |q_fB|}{2\pi T} \int_{-\infty}^{+\infty}\frac{dk_3}{2\pi} n_F(k_3)  (1 - n_F(k_3)) \nn 
&=&  \sum_f \frac{g^2 |q_fB|}{4\pi^2}   
\label{mDq} 
\eea 
where $n_F(k_3) = (e^{\beta(k_3 - \mu)}  + 1)^{-1}, ~~\beta = T^{-1}$, is the Fermi-Dirac distribution function. The expression for $m_{D,q}^2$ is different from the obtained for dense QCD matter in Ref. \cite{Andersen:2002jz}, which has a $\mu^2$ type functional dependence. In the $T \to 0$ limit, $m_{D,g}^2$ has no contribution, whereas $m_{D,q}^2$ is evaluated by replacing the Fermi-Dirac distribution function with a Heaviside step function. A difference to be noted is that in a strong magnetic field, we work in $(1+1)$ dimensions, whereas in the absence of a magnetic field, as studied in Ref. \cite{Andersen:2002jz}, we work in (3+1) dimensions. In Ref. \cite{Bandyopadhyay:2016fyd}, it has been shown that in a strong magnetic field limit, the Debye mass has no medium dependence. In contrast, for massive quarks, there is a separation of energy scales in the system, which induces a medium dependence in the Debye mass. Similar results have been obtained in Refs. \cite{Karmakar:2018aig,Khan:2021syq} as well. But if one considers higher Landau levels, the Debye mass at $\mu \neq 0$ and $T = 0$ is given by~\cite{Huang:2023hlk} 
\bea
m_{D,q}^2(\mu, q_fB) = \frac{g^2 \mu}{(2\pi)^2}\sum_f \sum_{l = 0}^{l_\text{max}} \frac{ |q_fB| (2 - \delta_{0,l}) }{ \sqrt{\mu^2  - 2l|q_fB|  } } ,
\label{mDq-l}
\eea
showing that there is a $\mu$ dependence for higher Landau levels.   
The above expression at $l = 0$ reduces to the result given by Eq. (\ref{mDq}). 
Focussing on the expression of the form factors $a(p_0, p_3)$ and $b(p_0, p_3)$ as calculated in Ref. \cite{Karmakar:2018aig}, the final expressions of the form factors for the dense fermions are given by 
\bea
a(p_0, p_3) & \approx &   \sum_f 2 e^{-p_\perp^2/q_fB} \frac{\sqrt{\overline{n}^2}}{\sqrt{\overline{u}^2}} \bigg(\frac{g^2 |q_fB|}{4\pi^2} \bigg)\frac{p_0 p_3}{p_0^2 - p_3^2} \nn
b(p_0, p_3) & \approx & -\sum_f \frac{g^2 |q_fB|}{4\pi^2 \overline{u}^2} e^{-p_\perp^2/2q_fB} \frac{p_3^2}{p_0^2 - p_3^2} = -d(p_0, p_3)\nn
c(p_0, p_3) &=& 0, 
\label{FG-4}
\eea
where $\overline{n}^2 = -p_\perp^2 /p^2$ and $\overline{u}^2 = -p^2/P^2$. 
Clearly, the ideal part of the gluon-free energy vanishes in the $T \to 0$ limit. The only contribution to the corrections to free energy is given by~\cite{Karmakar:2019tdp} 
\bea
F_g' = -(N_c^2 - 1)\sumintob \bigg[ \frac{b + c + d}{2P^2} + \frac{b^2 + c^2 + d^2}{4P^4} \bigg].
\label{FG5}
\eea 
Using the expressions of the form factors given by Eq. (\ref{FG-4}), the expressions in the brackets of $F_g'$ are given by
\bea
\frac{b + c + d}{P^2} &=& 0  \quad , \quad \quad 
\frac{b^2 + c^2 + d^2}{4P^4} = \frac{1}{2}\sum_{f_1,f_2} \bigg(\frac{g^2 B}{4\pi^{2}}\bigg)^2 q_{f_1} q_{f_2} \sumintob  e^{-\frac{p_\perp^2}{2B}\big(\frac{1}{q_{f_1}} + \frac{1}{q_{f_2}}\big)} \frac{p_3^4}{p^4(p_0^2 - p_3^2)^2}. 
\label{FG6} 
\eea 
The zero temperature limit of the sum integral in Eq. (\ref{FG6}) is calculated as 
\bea
\displaystyle{\lim_{T \to 0}}\sumintob \,  e^{-\frac{p_\perp^2}{2B}\big(\frac{1}{q_{f_1}} + \frac{1}{q_{f_2}}\big)} \, \frac{p_3^4}{p^4(p_0^2 - p_3^2)^2} 
 &=& \displaystyle{\lim_{T \to 0}}  -\bigg(\frac{e^{\gamma_E}\Lambda^2}{4\pi}\bigg)^\epsilon \int \frac{d^{3-2\epsilon}p}{(2\pi)^3} e^{-p_\perp^2/2B\big(\frac{1}{q_{f_1}} + \frac{1}{q_{f_2}}\big)} \frac{p_3}{2p^4}\bigg[ \beta \frac{\partial}{\partial \beta}  - 1\bigg] n_B(p_3)  \nn 
&=& \frac{1}{(4\pi)^2}\bigg[ \frac{1}{2\epsilon} + \frac{\ln 4}{2}  + \gamma_E \bigg] + \text{higher orders in } \frac{1}{q_fB}, 
\label{FG7}  
\eea 
where $\beta = T^{-1}$. On substituting Eq. (\ref{FG7}) in $F_g^{\prime}$ we get 
\bea
F_g^{\prime} &=& -\frac{(N_c^2 - 1)}{2}\sum_{f_1, f_2} q_{f_1} q_{f_2}\bigg(\frac{g^2 B}{4\pi^{2}}\bigg)^2 \frac{1}{(4\pi)^2}\bigg[ \frac{1}{2\epsilon} + \frac{\ln 4}{2}  + \gamma_E \bigg] ,
\label{FG8}
\eea 
where we see that there are $\mathcal{O}(\epsilon^{-1})$ terms present in $F_g$, devoid of any medium dependence, that renders the free energy divergent. In order to remove these divergences, we add a counterterm free energy $F_g^\text{ct}$ given by 
\bea
F_g^\text{ct} = \frac{(N_c^2 - 1)}{4\epsilon}\sum_{f_1, f_2} q_{f_1} q_{f_2} \, \bigg(\frac{g^2 B}{4\pi^{2}}\bigg)^2 \frac{1}{(4\pi)^2},
\label{FG9}  
\eea 
to $F_g$, which renormalizes the free energy. Thus, the final expression of the gluon free energy $F_g$ is given by
\bea
F_g = F_g^{\prime} + F_g^\text{ct},
\label{FG10}
\eea 
where the $\mathcal{O}(\epsilon^{-1})$ divergences have been removed via $F_g^\text{ct}$. We note that $F_g$ is independent of the $\mu$, unlike the quark free energy. This is attributed to the quark and antiquark propagators in the gluon self-energy diagram considered here, which exactly cancels the effect of $\mu$. 
\subsection{Quark number susceptibility and speed of sound}
\label{cs}
In the previous section, we computed the ideal pressure and HDLpt corrected pressure, which will be utilized to estimate the second-order QNS and speed of sound. From the thermodynamic relation
\bea
\varepsilon - F =  \mu n + e \mathcal{M}\cdot B~,
\label{therm-1}
\eea
we obtain the energy density $\varepsilon$ as a sum of the ideal part along with the HDLpt corrections. Here $F$ is the total free energy, and $n$ is the number density. From Eq. (\ref{therm-1}) number density can be obtained as 
\bea
n = -\frac{\partial F}{\partial \mu}~.
\eea 
Now, let us define the QNS, which is a measure of the response of conserved quark number density $n$, with the infinitesimal change in quark chemical potentials $\mu + \delta \mu$. In QCD thermodynamics, it is generally defined as the second-order derivative of the pressure ${P}$ with respect to quark chemical potential $\mu$. However, due to the presence of anisotropy in the pressure, one can define two different second-order QNS, namely along the longitudinal and transverse directions in a strongly magnetized medium. Thus, the longitudinal second-order QNS is defined as
\begin{equation}
\chi^{2}_{_L}=\left.\frac{\partial n}{\partial \mu}\right|_{\mu \rightarrow 0}=\left.\frac{\partial^2 \mathcal{P}_{L}}{\partial^2 \mu}\right|_{\mu \rightarrow 0},
\end{equation}
while the transverse second-order QNS can be obtained as
\begin{equation}
\chi^{2}_{_\perp}=\left.\frac{\partial n}{\partial \mu}\right|_{\mu \rightarrow 0}=\left.\frac{\partial^2 \mathcal{P}_{\perp}}{\partial^2 \mu}\right|_{\mu \rightarrow 0}
\end{equation}
whereas the second-order QNS for the ideal dense magnetized system can be obtained from the Eq.~\eqref{F_Ideal}. Also, in an external magnetic field, the speed of sound $c_s$ breaks into two components viz. $c_{s,\parallel}$ and $c_{s,\perp}$. These can be computed using the relation
\bea
(c_{s,j})^2 = \left.\frac{\partial  P_j}{\partial \varepsilon}\right|_{j = \parallel, \perp},
\eea 
where $P_j$ are the components of pressure.
\section{Results and Discussions}
\label{Res}
In this section, we discuss the results of the anisotropic pressure, magnetization, second-order quark number susceptibility, and speed of sound subjected to the strong magnetic field. For this, we note that in a strong magnetic field, the strong coupling $\alpha_s = \frac{g^2}{4\pi}$ depends not only on chemical potential but also on the magnetic field. The expression of the one-loop running coupling constant in a strong magnetic field \cite{Ayala:2018wux} is given by 
\bea
\alpha_s(\Lambda^2, |q_fB|) = \frac{ \alpha_s(\Lambda^2)  }{1 + b_1\alpha_s(\Lambda^2)\ln\big( \frac{\Lambda^2}{\Lambda^2 + |q_fB|} \big)    },~~\text{where~~} \alpha_s(\Lambda^2) = \frac{1}{ b_1\ln\big( \Lambda^2 / \Lambda^2_{ \overline{\text{MS}} } \big)   },
\label{res-1}
\eea    
where $b_1 = (11N_c - 2N_f)/12\pi$, $\Lambda_{ \overline{ \text{MS} } } = 176$ MeV \cite{Karmakar:2020mnj} and the renormalization scale $\Lambda = 2\mu$ \cite{Fraga:2023lzn}. Since we are interested in the thermomagnetic correction here, we drop the vacuum contribution $-B^2/2$ in our results. The total free energy of the system is the sum of the quark free energy $F_f$ given by Eq. (\ref{F_HDL}) and gluon free energy $F_g$ given by Eq. (\ref{FG10}). The qualitative features of this work have been shown in the plots by analyzing their behavior with chemical potential and magnetic field. 
\begin{figure}[h]
	\centering
	\includegraphics[scale = 0.3]{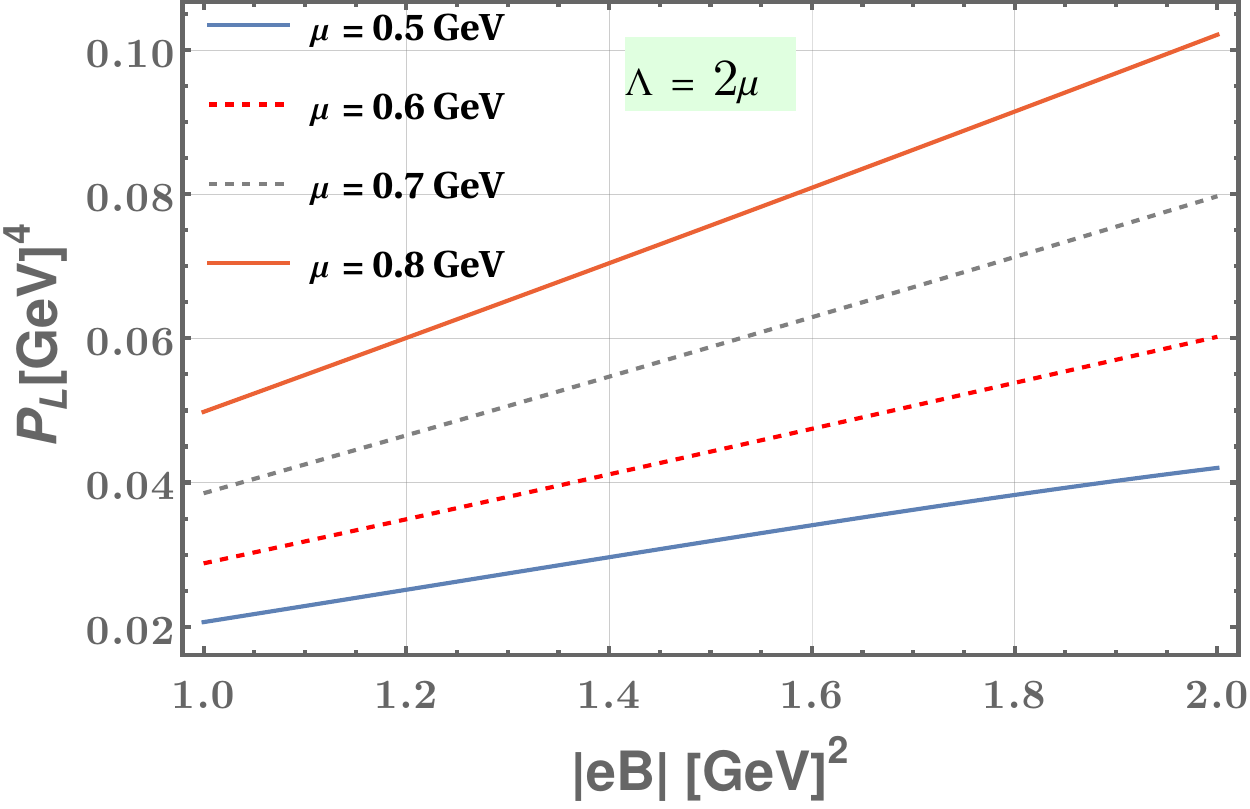}
	\includegraphics[scale = 0.3]{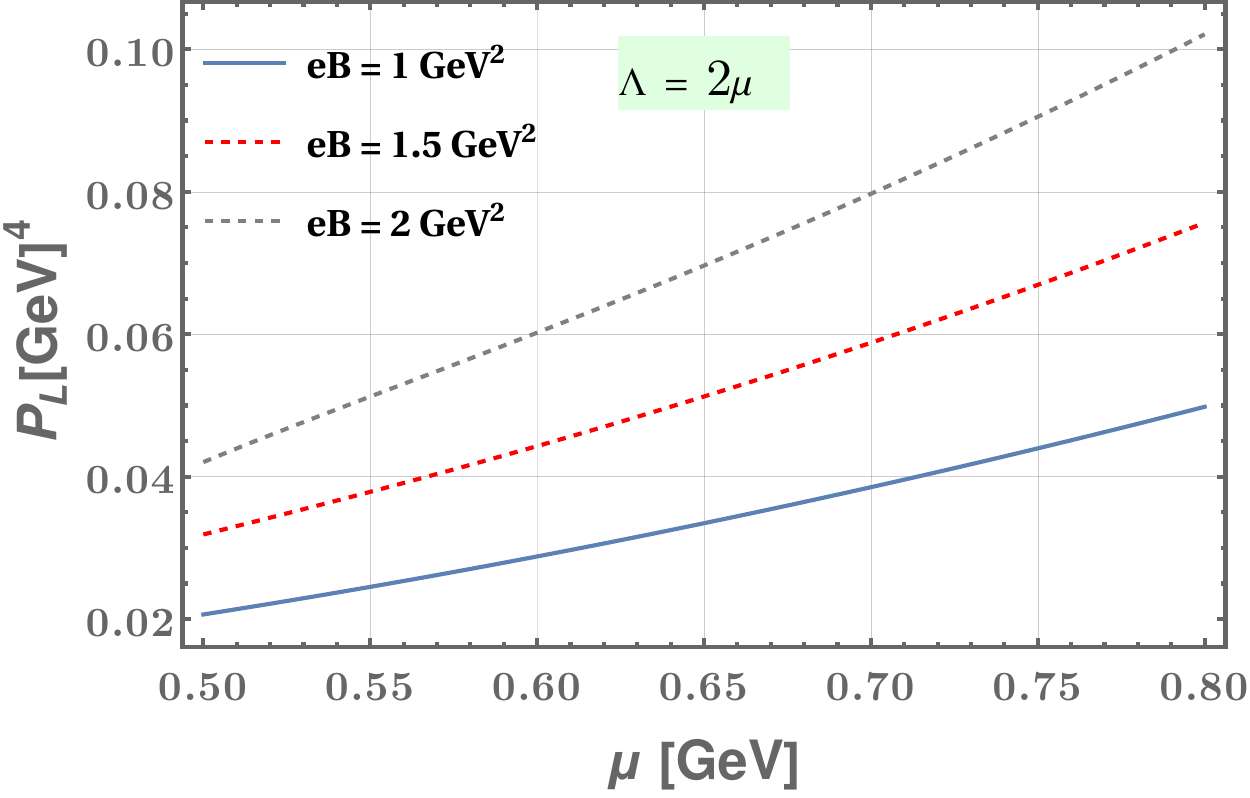}
	\caption{Plots of the longitudinal pressure with magnetic field and chemical potential. (a) The left panel shows the variation with magnetic field for different chemical potentials, and (b) the right panel shows the variation with chemical potential for different magnetic fields. }
	\label{fig-1}
\end{figure}

\begin{figure}[h]
	\centering
	\includegraphics[scale = 0.3]{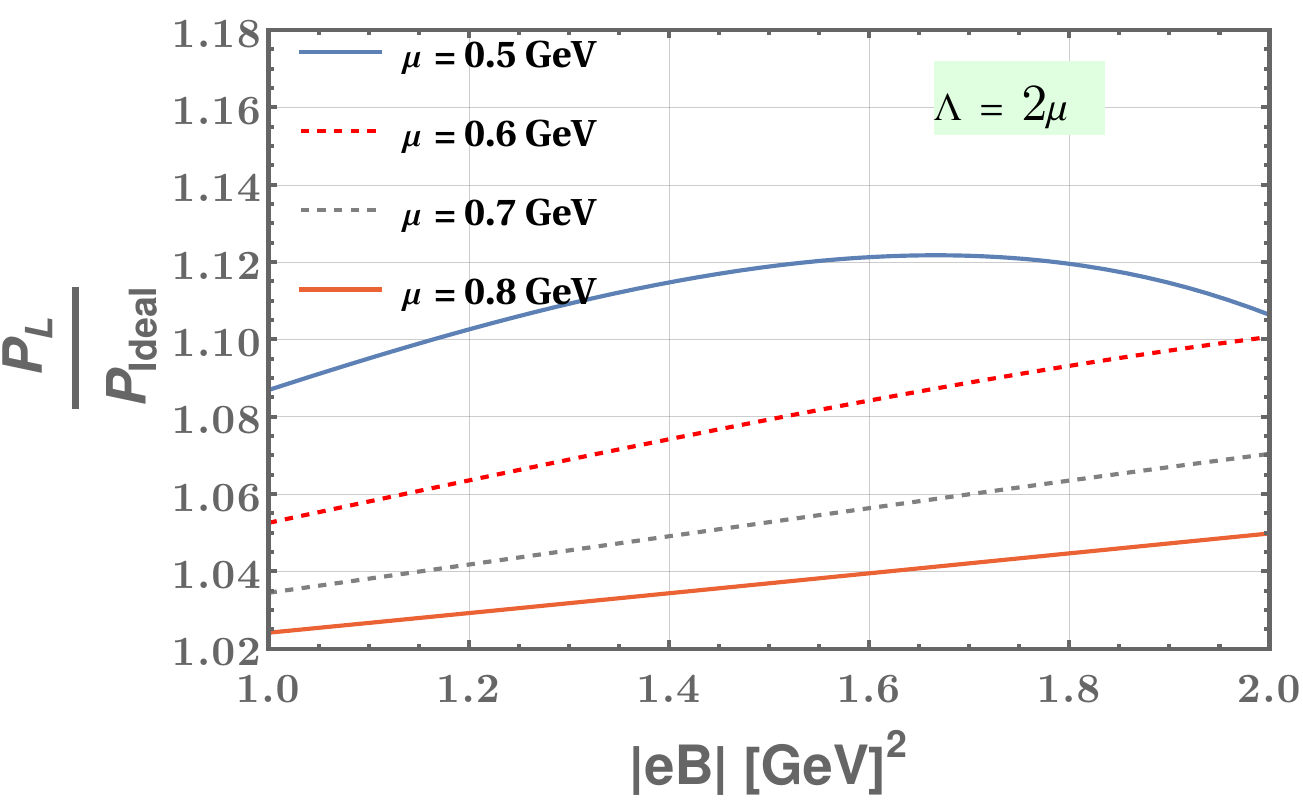}
	\includegraphics[scale = 0.3]{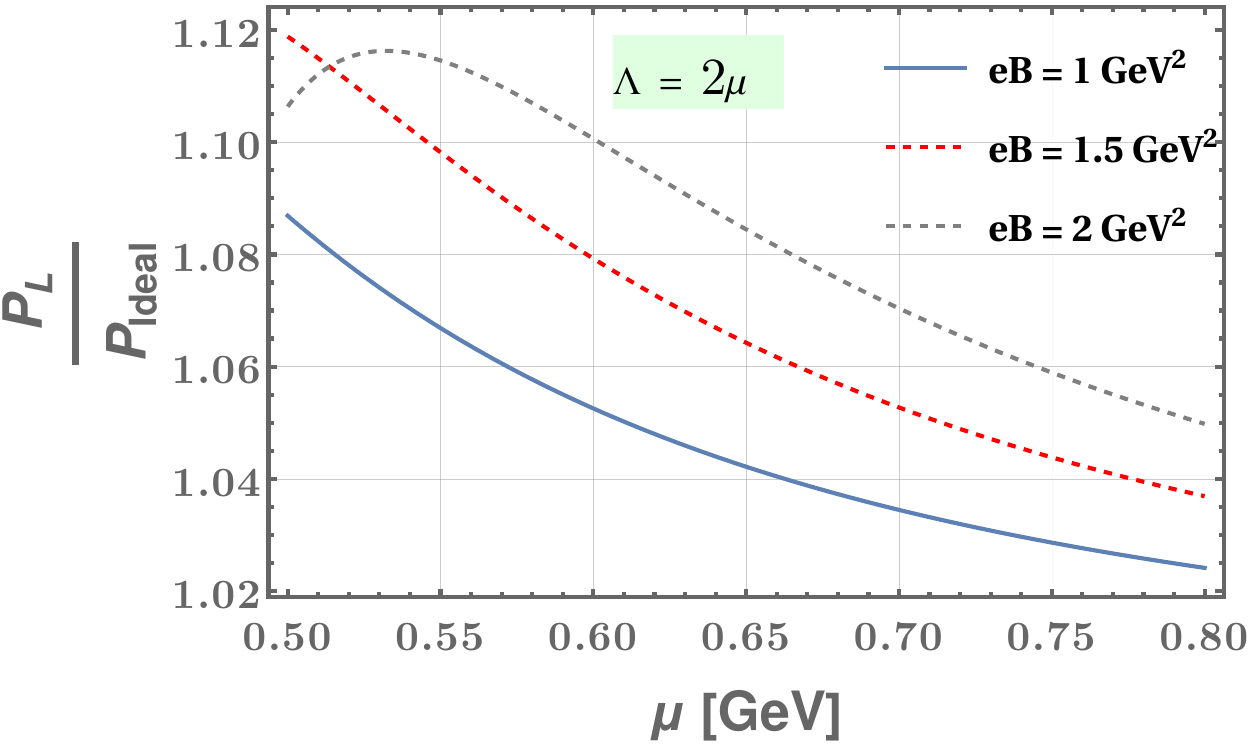}
	\caption{Plots of the ratio of the longitudinal to ideal pressure with magnetic field and chemical potential. (a) The left panel shows the variation with magnetic field for different chemical potentials, and (b) the right panel shows the variation with chemical potential for different magnetic fields.}
	\label{fig-2}
\end{figure}
The behavior of the longitudinal pressure has been shown in Fig.~\ref{fig-1} with magnetic field and chemical potential on the left and right panel, respectively. Since we are dealing with a strongly magnetized medium, the range of magnetic field has been taken between 1 to 2 GeV$^2$, which satisfies the LLL condition. We have chosen the chemical potential between 0.5 to 0.8 GeV, where the upper limit has been chosen in accordance with the value expected in neutron stars \cite{Fraga:2023lzn}. It is observed that the longitudinal pressure increases with the magnetic field and chemical potential. This behavior is expected in quark matter when the magnetic field and density are extremely high, leading to an amplification in thermodynamic and transport properties. A comparison with ideal pressure is plotted in terms of the ratio between the longitudinal and ideal values in Fig.~\ref{fig-2}. The plots on the left panel show that the ratio converges toward unity or the HDLpt corrected pressure approaches the ideal value at a low magnetic field and diverges away from unity as the magnetic field increases. The plots on the right panel show that at extremely high density, the plots converge to unity with a decreasing trend. This behavior is expected when temperature and densities become large. This decreasing behavior can be attributed to the fact that the HDLpt corrections have a functional dependence on $\mu$ of the form $\frac{N_cq_fB}{4\pi^2}\big[ \mathcal{O}(\mu^{-2n})  \big]$, where $n = 1, 3$. The behavior of these plots is reminiscent of what has been obtained by Karmakar {\it et al.} in Ref. \cite{Karmakar:2019tdp} at high temperatures using HTLpt. From here, we see that temperature and chemical potential are complementary to each other in their asymptotic limits, and their effect is highly pronounced in the high $T$ and high $\mu$ regimes, where they attain ideal values in the HTLpt/HDLpt framework. Apart from that, for a given strength of the magnetic field, an $\mathcal{O}(B)$ dependence is seen for the ideal case. This functional dependence is of higher powers of $B$ for the HDL corrections coming from the one-loop diagrams. This leads the longitudinal pressure to increase with a magnetic field. On comparing our results with that obtained in Ref. \cite{Fraga:2023lzn} via renormalization group optimized perturbation theory (RGOPT), we find that the behavior and the order of magnitude of the longitudinal pressure are the same. The results are not directly comparable with other methods, such as LQCD, since LQCD does not work for finite or high $\mu$. \par
\begin{figure}[h]
	\centering
	\includegraphics[scale = 0.3]{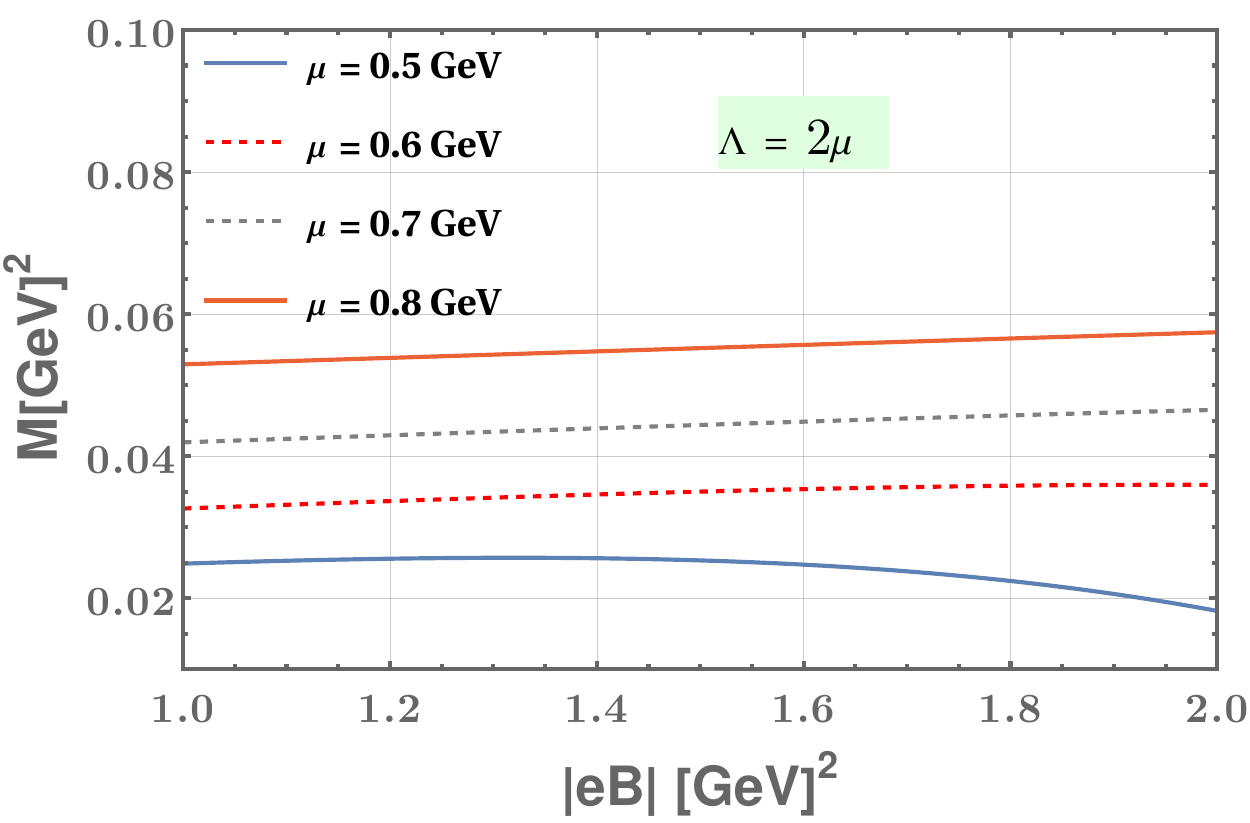}
	\includegraphics[scale = 0.3]{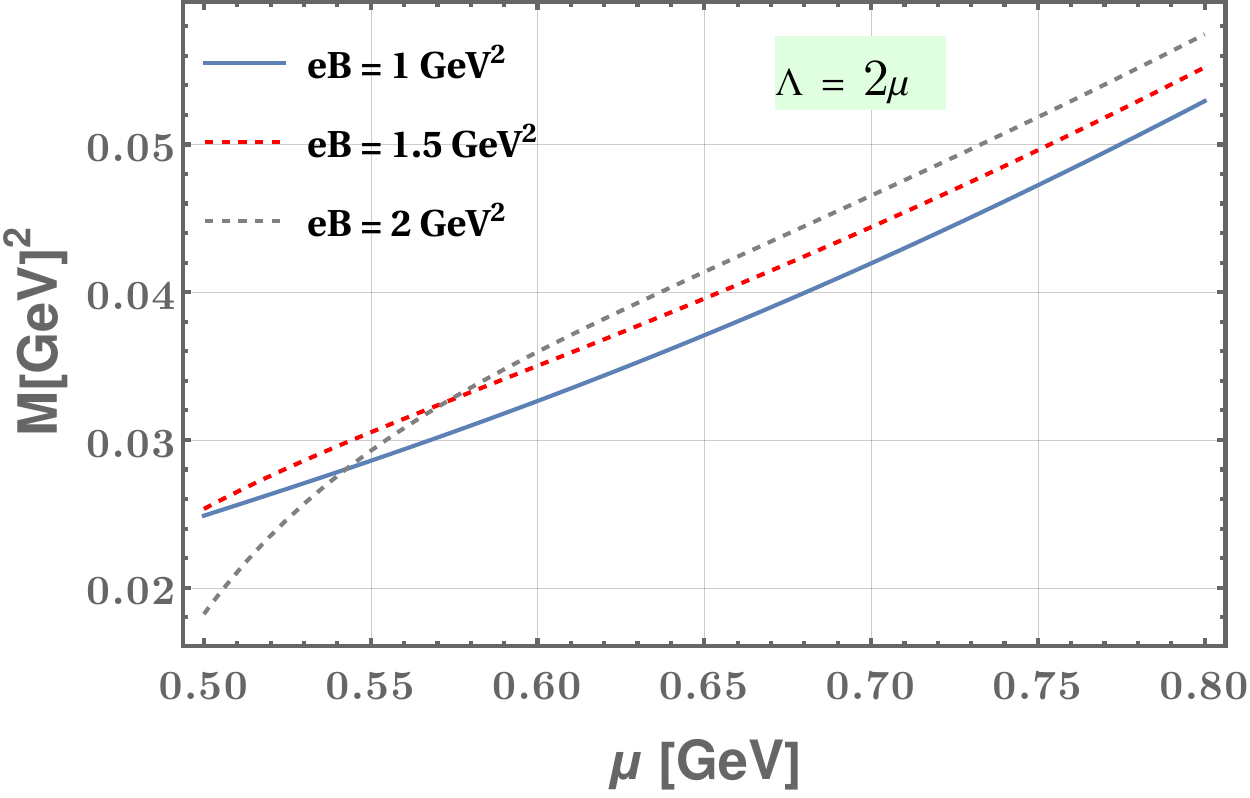}
	\caption{Plots of the magnetization with magnetic field and chemical potential. (a) The left panel shows the variation with magnetic field for different chemical potentials, and (b) the right panel shows the variation with chemical potential for different magnetic fields. }
	\label{fig-3}
\end{figure}
\begin{figure}[h]
	\centering
	\includegraphics[scale = 0.3]{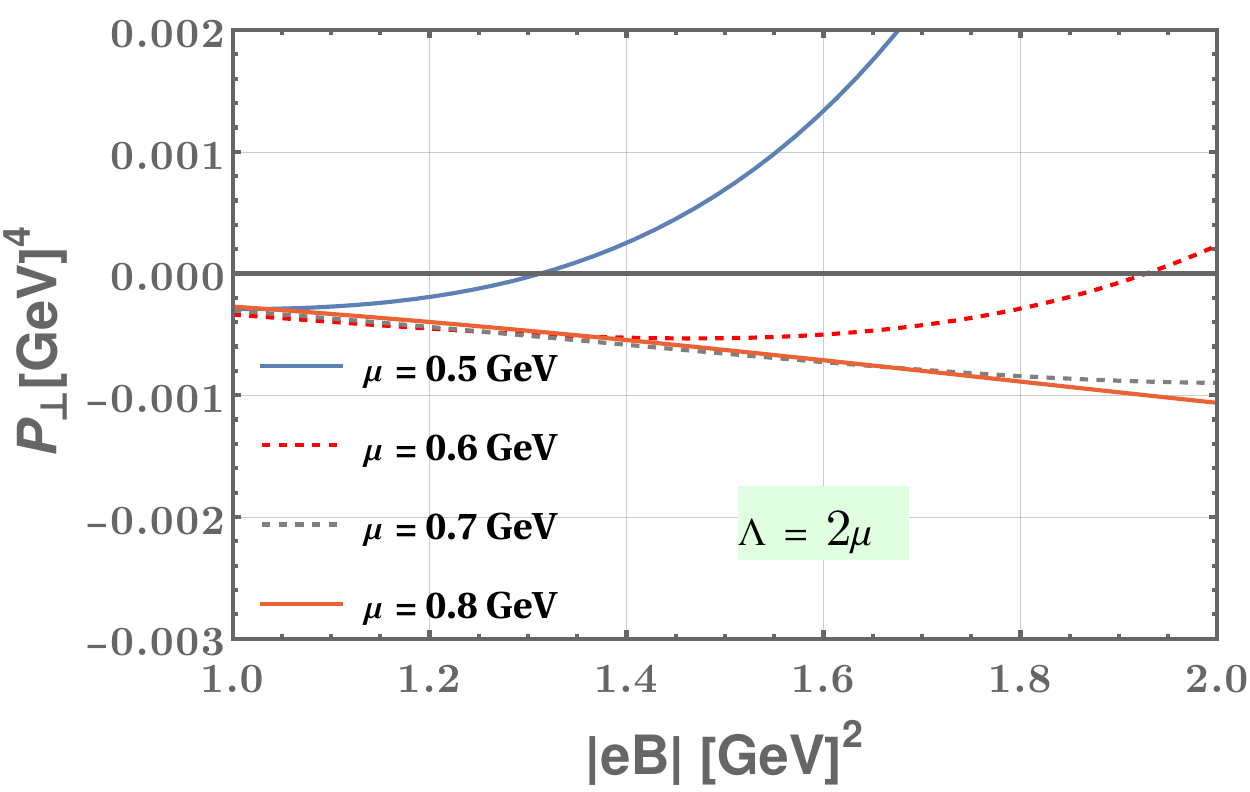}
	\includegraphics[scale = 0.3]{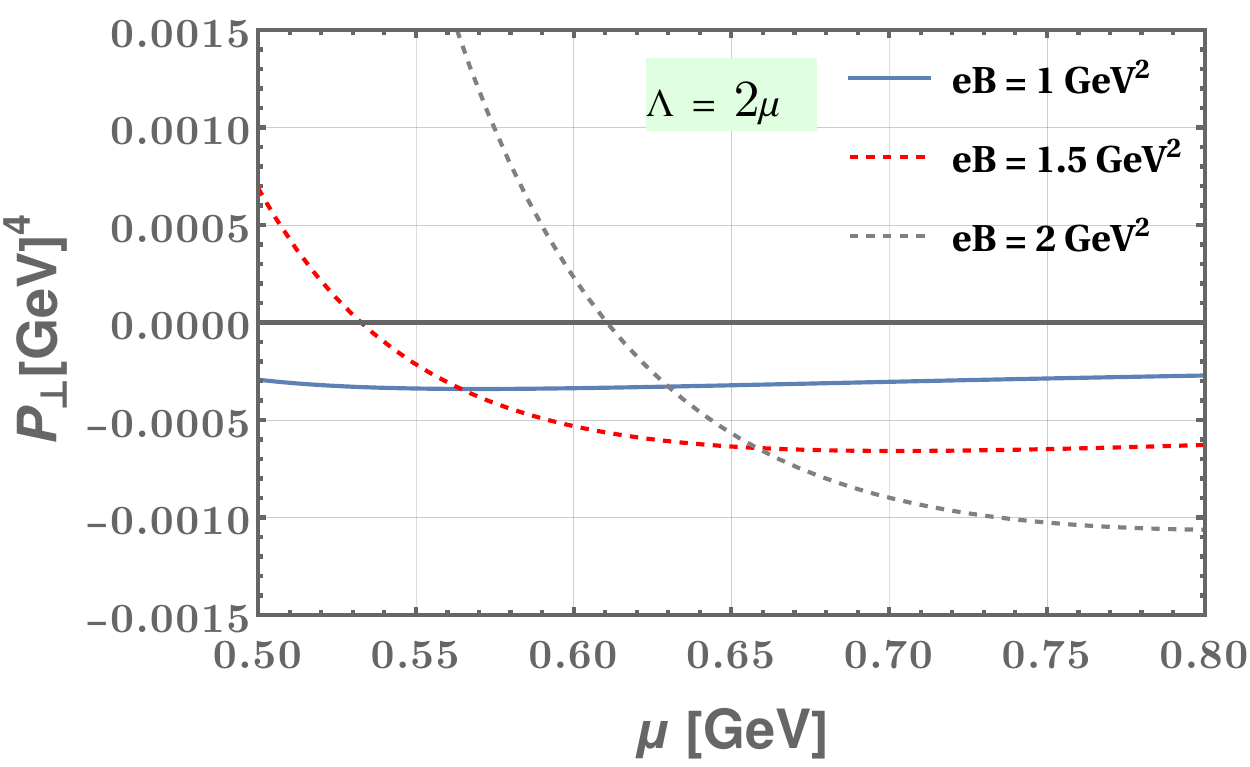}
	\caption{Plots of the transverse pressure with magnetic field and chemical potential. (a) The left panel shows the variation with magnetic field for different chemical potentials, and (b) the right panel shows the variation with chemical potential for different magnetic fields. }
	\label{fig-4}
\end{figure}

The behavior of magnetization and a transverse component of pressure with magnetic field and quark chemical potential is shown in Fig.~\ref{fig-3} and Fig.~\ref{fig-4}, respectively. As shown in Fig.~\ref{fig-3}, magnetization is found to increase with magnetic field and chemical potential. This shows that strongly magnetized quark matter is paramagnetic and, in turn, affects the transverse component of pressure. 
Nevertheless, as we computed the magnetization, it became straightforward to compute the transverse pressure, which has been shown in Fig.~\ref{fig-4}. The plots show that for the strengths of the magnetic field and chemical potential considered in this work, the transverse pressure attains negative values and eventually attains positive values. It should be noted that these values are of the order of 10$^{-4}$ to 10$^{-3}$, which are very close to zero. The appearance of these nonzero values can be attributed to the functional dependence of $P_L$ and $\mathcal{M}$ on $\mu$ and $eB$. 

\begin{figure}[h]
	\centering
	\includegraphics[scale = 0.28]{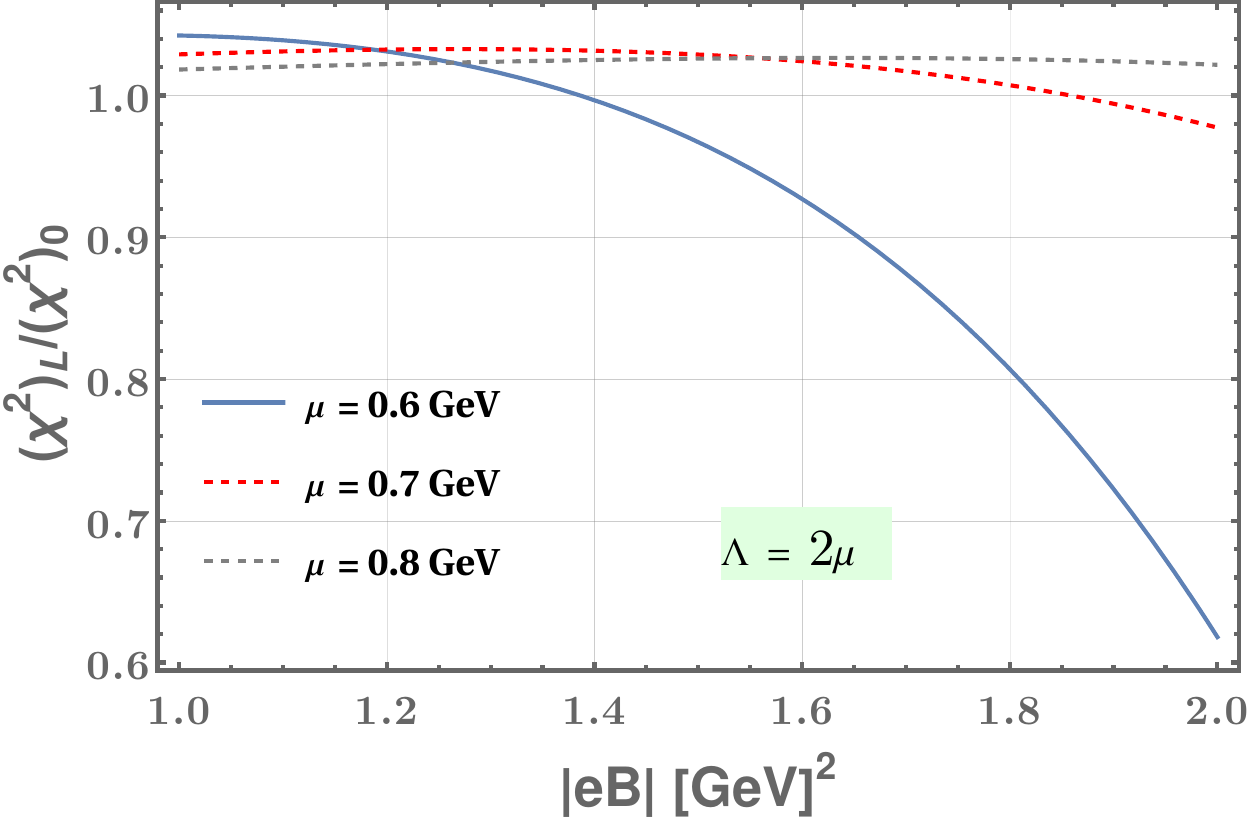}
	\includegraphics[scale = 0.28]{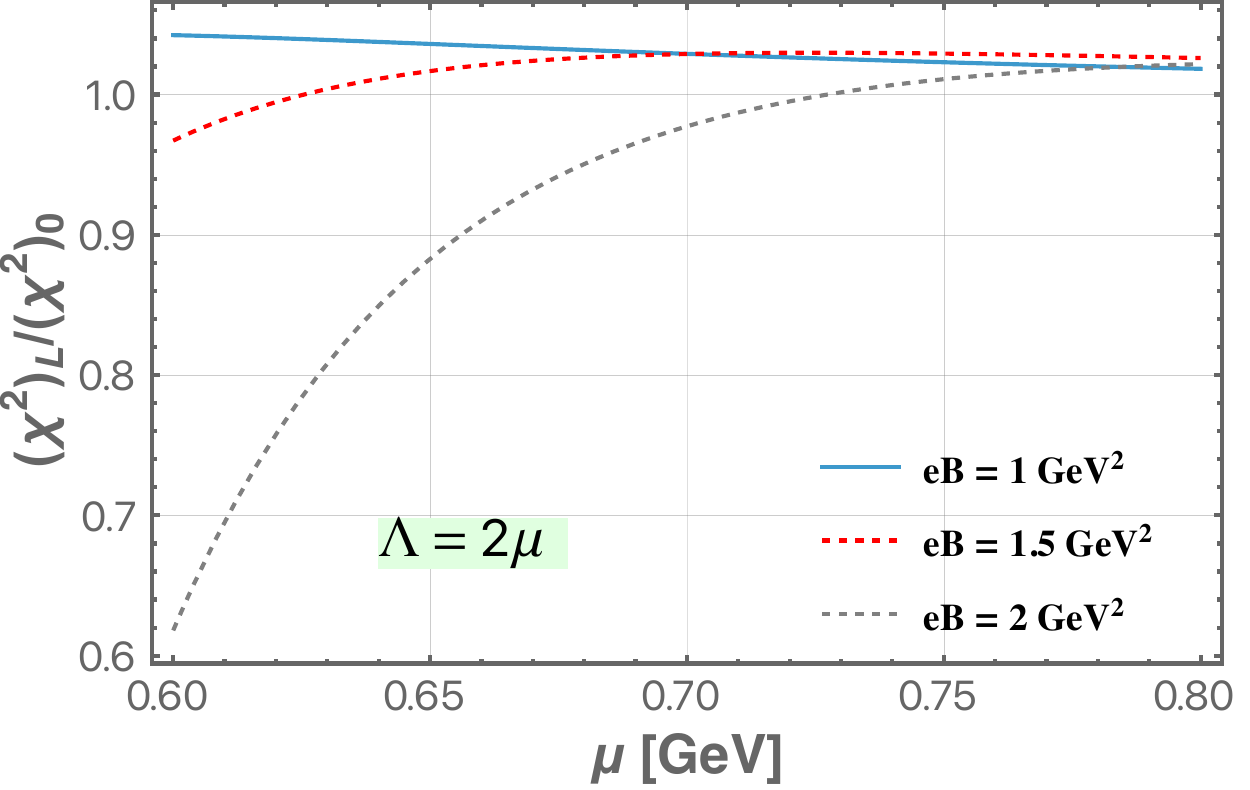}
	\caption{Variation of longitudinal second-order QNS scaled with a free field value in the presence of strongly magnetized medium with quark chemical potential (left) and with the magnetic field (right).}
	\label{fig-5}
\end{figure}

\begin{figure}[h]
	\centering
	\includegraphics[scale = 0.28]{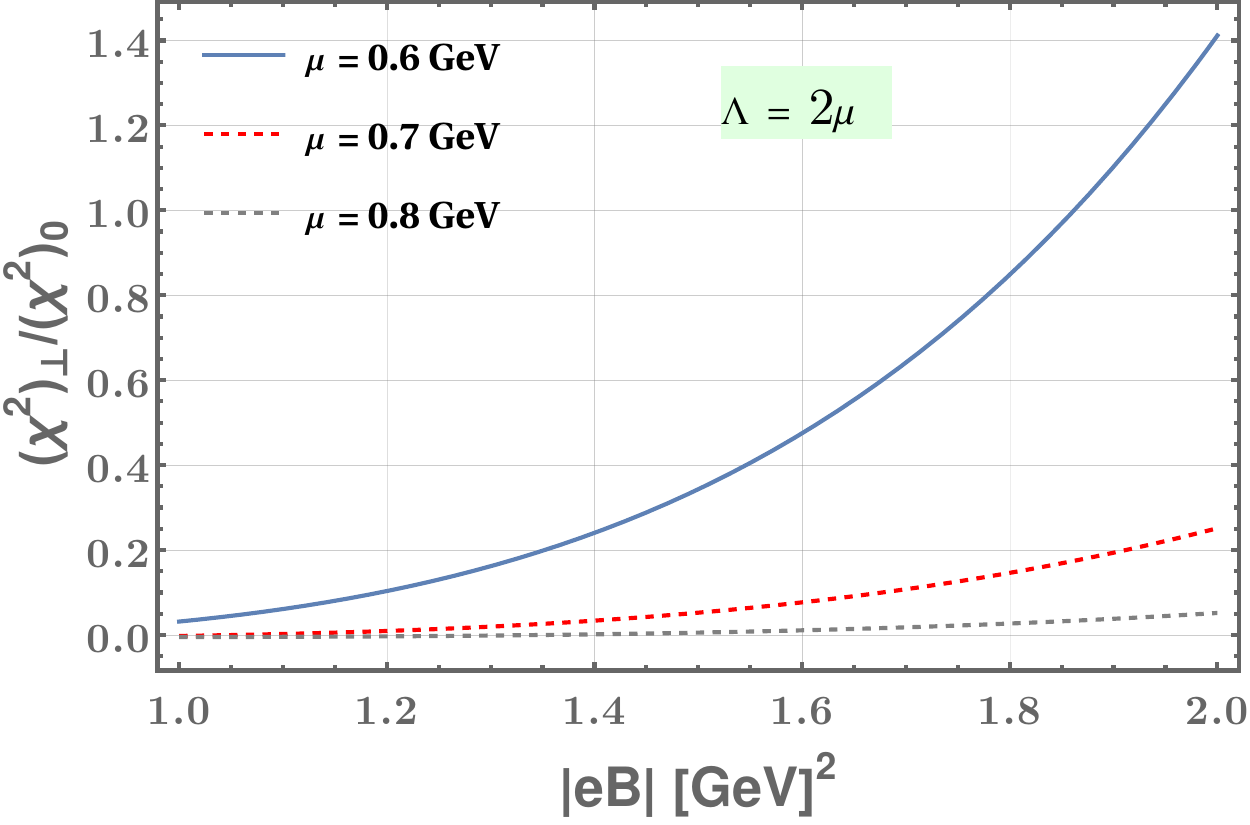}
	\includegraphics[scale = 0.28]{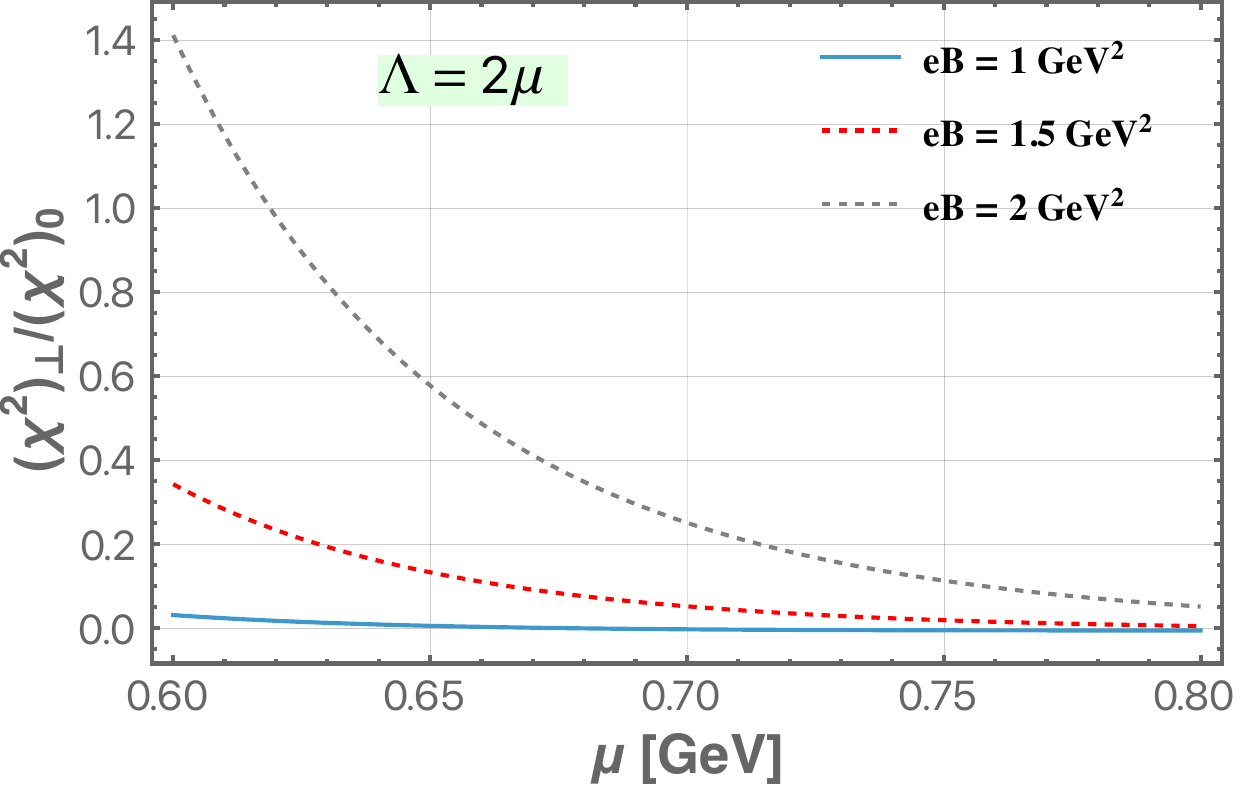}
	\caption{Variation of transverse second-order QNS scaled with a free field value in the presence of strongly magnetized medium with quark chemical potential (left) and with magnetic field (right).}
	\label{fig-6}
\end{figure}

\begin{figure}
	\centering
	\includegraphics[scale = 0.28]{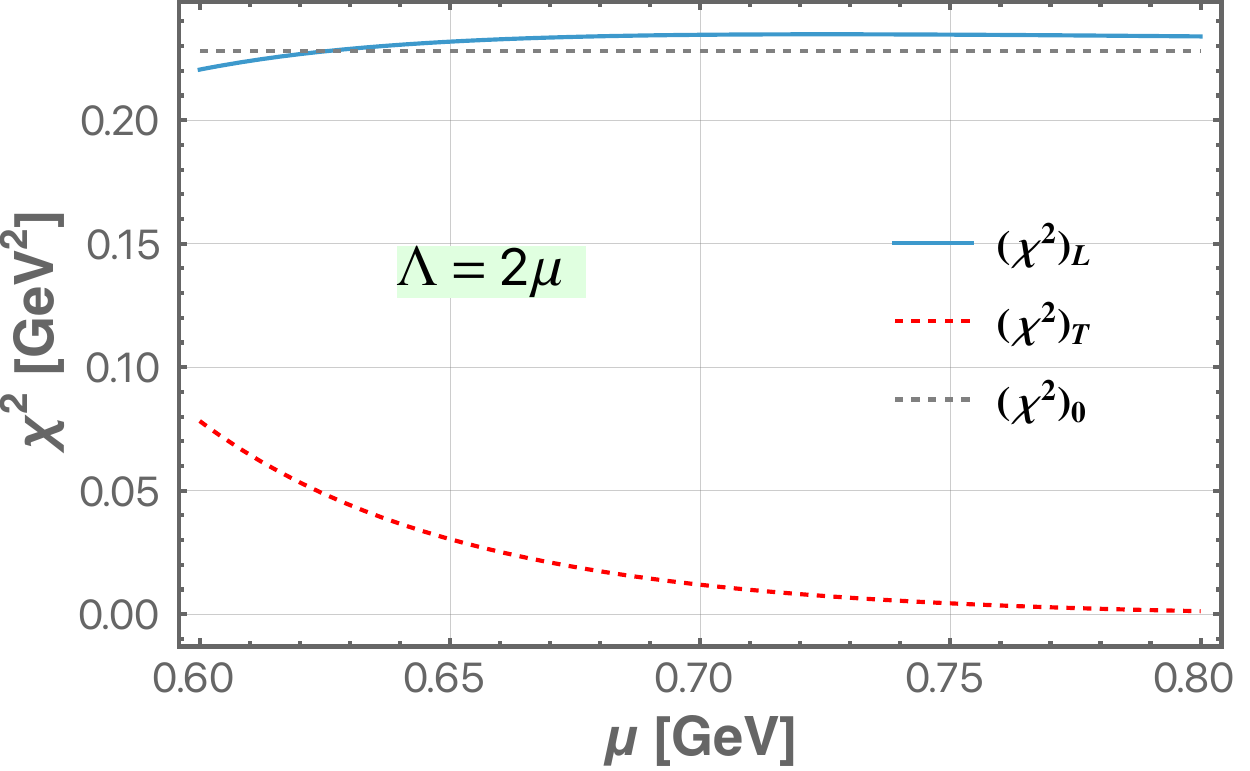}
	\caption{Variation of longitudinal and transverse second-order QNS with quark chemical potential.}
	\label{fig-7}
\end{figure}

We have plotted the behavior of second-order quark number susceptibility, namely longitudinal second-order QNS, and transverse second-order QNS, scaled with ideal QNS,  with magnetic field and quark chemical potential in Fig.~\ref{fig-5} and Fig.~\ref{fig-6} respectively. It can be seen in Fig.~\ref{fig-5} (left panel) that longitudinal QNS decreases with the magnetic field, and the decrease is more prominent for the lower-density regime. In contrast, it has been observed in Fig.~\ref{fig-6} (left panel) that the transverse second-order QNS increases with the magnetic field, and again, the increase is more dominating for lower-density values. The longitudinal second-order QNS increases with quark chemical potential as shown in Fig.~\ref{fig-5} (right panel) and approaches the ideal value for high dense regime irrespective of magnetic field strength. In contrast, the transverse second-order QNS decreases, as can be seen in Fig.~\ref{fig-6} (right panel), indicating that the system shrinks in the transverse direction. Note that since the QNS measures the fluctuations in quark number density, it can be seen that the magnitude for longitudinal second-order QNS is more compared to transverse second-order QNS as shown in Fig.~\ref{fig-7} due to the dynamics of the governed system. Also, longitudinal QNS contributes more than ideal QNS since longitudinal QNS contribution comes from the magnetic field as well as through quark chemical potential. In contrast, an ideal QNS depends only on the strength of the magnetic field. \par   

\begin{figure}
	\centering
	\includegraphics[scale = 0.3]{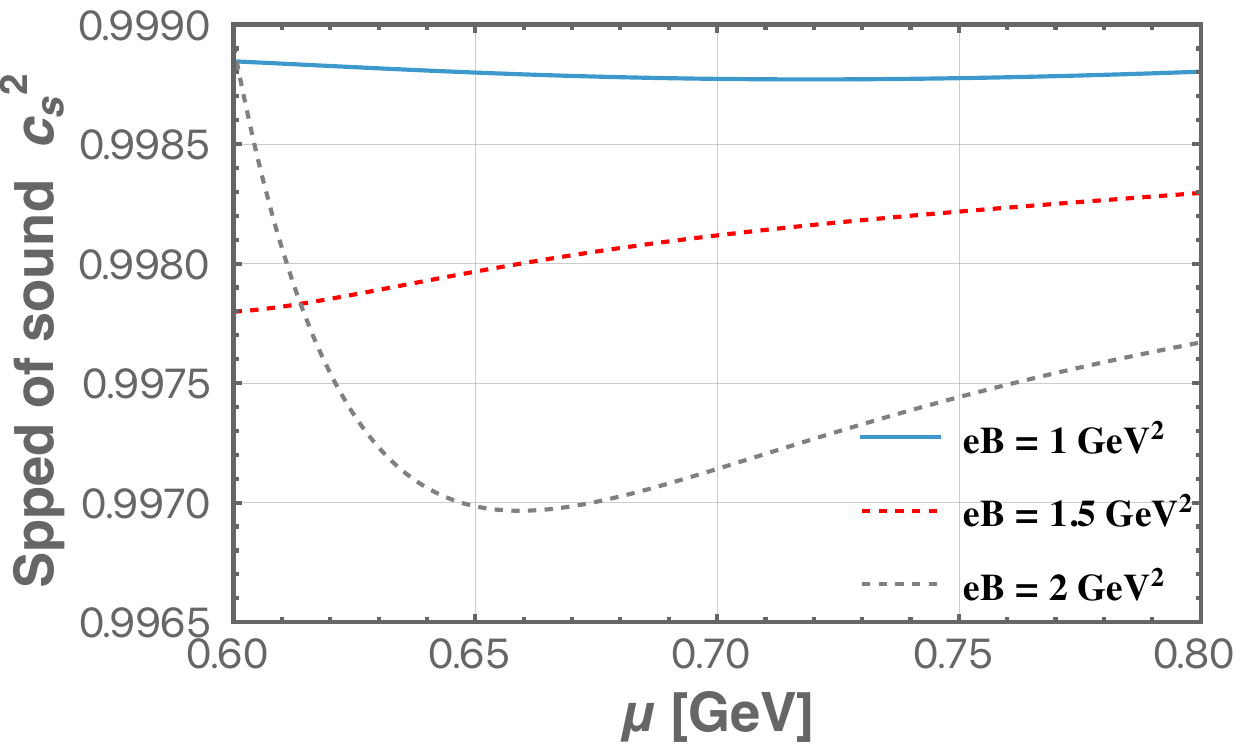}
	\caption{Plot of the longitudinal component of the square of the speed of sound with chemical potential for different magnetic fields. }
	\label{fig-8}
\end{figure}

In Fig.~\ref{fig-8}, we have shown the variation of $c_{s,\parallel}^2$, the longitudinal component of the square of the speed of sound, with $\mu$ for different magnetic fields. It is found that it approaches $c$, the speed of light, in the regime of high density, irrespective of the strength of the magnetic field, showing a universal behavior in this regime. Although the upper bound on $c_s^2$ is 1/3 in $(3+1)$ dimensions, however, at very high densities \cite{Zeldovich:1961sbr}, such as in astrophysics, this bound could be violated. Studies of high-density matter based on HDLpt in Ref. \cite{Fujimoto:2020tjc} show that $c_s^2 > 1/3$ for lower densities and approaches 1/3 for the higher density regime. Similar results have been reported in the Walecka model \cite{Kapusta:2007xjq} in dense nuclear matter, where the effective nucleon mass vanishes. This leads to the pressure being equal to the energy density at high density, and the speed of sound approaches the speed of light. A study of dense nuclear matter based on the Walecka model in the magnetic field in Ref. \cite{Mondal:2023baz} shows that the square of the speed of sound very clearly exceeds 1/3 and goes as high as 0.8. Concerned with the present study, we should note that in the absence of a magnetic field, one works in (3+1)$d$, whereas in LLL, the dynamics of the system are (1+1)$d$. Studies based on high-density matter in Refs. \cite{Ferrer:2010wz, Zeldovich:1961sbr}, report that the thermodynamic potential behaves as $\sim \mu^{d + 1}$, where $d$ is the number of spatial dimensions and from this observation. This leads to the speed of sound to vary as $c_s^2 \lesssim 1/d$. 
It is expected that the speed of sound should not exceed the speed of light, and this is ensured by the HDLpt corrections to the free energy, which, through a gradual addition of $\mathcal{O}(g^{2n})$ corrections to the longitudinal pressure and energy density, ensure that this condition is not violated. The results in Fig.~\ref{fig-8} verify this property where we see that the plots approach unity from below, thus obeying the condition $c_{s,\parallel} < c$. We have neglected the transverse component, i.e., $c_{s,\perp}^{2}$, owing to its vanishing values in a strong magnetic field.

\section{Conclusions and outlook}
\label{conc}
In this paper, we have studied the thermodynamics of strongly magnetized quark matter in the framework of HDLpt at one-loop. Due to the presence of a magnetic field, the thermodynamic quantities acquire a multicomponent structure relative to the direction of the magnetic field. Further, due to the strong magnetic field, it is observed that the longitudinal components have a sizeable contribution, as observed in the longitudinal pressure. The magnetization has been computed, and the plots show a paramagnetic nature. The behavior of longitudinal and transverse second-order QNS has been estimated for different strengths of magnetic field and quark chemical potentials. The longitudinal component of the square of the speed of sound has been computed, where the results show that it asymptotically approaches $c$ at high $\mu$. This behavior can be attributed to the number of spacetime dimensions available to the fermions. In the absence of an external magnetic field, the dynamics can be studied in (3+1)$d$, whereas it changes to (1+1)$d$ in a strong magnetic field. A further study of the trace anomaly in strong magnetic fields in dense scenarios will shed more light on the results of this work. We plan to address this in future work. The present work has been confined to one loop only. Therefore, the corrections up to $\mathcal{O}(g^4)$ have been considered only. An extension of HDLpt to higher loops will enable it to go beyond $\mathcal{O}(g^4)$ corrections.

\section*{ACKNOWLEDGEMENTS}
We thank Najmul Haque, Hiranmaya Mishra, Sukanya Mitra, Amaresh Jaiswal, Rajkumar Mondal, and Deeptak Biswas for the valuable discussions. Sumit would like to thank Fei Gao for the helpful discussion. S.A.K. is thankful to Integral University for providing the necessary facilities for research and assigning the manuscript communication number IU/R\& D/2025-MCN0003519.

\section*{DATA AVAILABILITY}
The data supporting this study’s findings are available within the article.

\appendix
\section{Dense sum integrals}
\label{App-A}
 
In Ref. \cite{Gorda:2022yex}, it was shown that for the dense case, the $T\to 0$ and $T = 0$ limit of the sum integrals at finite $T$ and $\mu$ are not equivalent. Evaluation of simple dense sum integrals showed that the $T \to 0$ limit is the correct approach for evaluating dense sum integrals. 
In the present work, we have sum integrals having different powers of temporal and spatial momenta in the numerator and different powers of four-momentum in the denominator. The general structure of fermionic sum integrals at finite $T$ and $\mu$ is given by
\bea
&&\sumintof  \frac{ p_0^{2\beta} p^{2\omega} }{ P^{2\alpha} }  \nn &=& \frac{ T ~~\Gamma(\alpha - \omega - d/2) \Gamma(d/2 + \omega) }{(4\pi)^{d/2} \Gamma(\alpha) \Gamma(d/2) (2\pi T)^{2\alpha - 2\beta - d - 2\omega}} \big\{ \zeta(2\alpha - 2\beta - 2\omega - d, 1/2 + i\mu/ (2\pi T)) +  \zeta(2\alpha - 2\beta - 2\omega - d, 1/2 - i\mu/ (2\pi T))  \big\} , \nn 
\label{App-1}
\eea 
where $\zeta(z, q) = \sum_{n = 0}^{\infty} \frac{1}{(n + q)^z}$ is the Hurwitz zeta function. For our case, we need the $T \to 0$ limit of Eq. (\ref{App-1}), which can be viewed as the dense version of Eq. (\ref{App-1}). Following the results of Ref. \cite{Gorda:2022yex}, the dimensionally regularized form of the dense sum integrals is given by
\bea  
\hspace{-0.55cm}\mathcal{I}_{\alpha\beta\omega}(\mu) =  \displaystyle{\lim_{T \to 0}}\sumintof  \frac{ p_0^{2\beta} p^{2\omega} }{ P^{2\alpha} }  
= \bigg( \frac{e^{\gamma_E} \Lambda^2}{4\pi}  \bigg)^\epsilon \frac{i\mu}{2\pi}\frac{ \Gamma(\alpha - \omega - d/2) \Gamma(d/2 + \omega) (i\mu)^{d + 2\omega - 2\alpha + 2\beta} }{ (4\pi)^{d/2} \Gamma(\alpha) \Gamma(d/2) \big( 1 + d + 2\omega - 2\alpha + 2\beta  \big)  } \Big\{ \big( -1\big)^{d + 2\omega - 2\alpha + 2\beta} -1 \Big\}. 
\label{SI-2}
\eea 
Using Eq. (\ref{SI-2}), the various sum integrals appearing in HDLpt corrected free energy of quarks in Eq. (\ref{I_abc}) comes out to be
\bea
&&\mathcal{I}_{630}(\mu) = \bigg(\frac{\Lambda}{4\pi \mu}\bigg)^{2\epsilon}\frac{63}{4\pi\mu^4}\bigg[\frac{1}{512} + \frac{-1 + \log 256 + 4\log\pi + 2\psi(0,11/2)}{1024}\epsilon  \bigg] + \mathcal{O}(\epsilon^2) \nn
&&\mathcal{I}_{621}(\mu) = \bigg(\frac{\Lambda}{4\pi \mu}\bigg)^{2\epsilon}  \frac{1}{4\pi\mu^4} \bigg[ \frac{7}{512} + \frac{179 - 210\gamma_E + 420 \log 2\pi}{15360}\epsilon  \bigg] + \mathcal{O}(\epsilon^2)  \nn
&&\mathcal{I}_{612}(\mu) = \bigg(\frac{\Lambda}{4\pi \mu}\bigg)^{2\epsilon} \frac{1}{4\pi\mu^4}\bigg[\frac{3}{512} - \frac{3(1 + 10\gamma_E - 20\log 2\pi)}{5120}\epsilon \bigg] + \mathcal{O}(\epsilon^2) \nn
&&\mathcal{I}_{603}(\mu) = \bigg(\frac{\Lambda}{4\pi \mu}\bigg)^{2\epsilon} \frac{3}{4\pi\mu^4} \bigg[ \frac{1}{512} - \frac{ 9 - 10\log 4 }{512}\epsilon  \bigg] + \mathcal{O}(\epsilon^2)   \nn 
&&\mathcal{I}_{210}(\mu) = \bigg(\frac{\Lambda}{4\pi \mu}\bigg)^{2\epsilon} \frac{1}{4\pi}\bigg[ \frac{1}{2\epsilon} + 1 - \frac{\gamma_E}{2} + \log 2\pi  \bigg] + \mathcal{O}(\epsilon) \nn
&&\mathcal{I}_{201}(\mu) =  \bigg(\frac{\Lambda}{4\pi \mu}\bigg)^{2\epsilon}\frac{1}{4\pi }\bigg[\frac{1}{2\epsilon}  -1 - \frac{\gamma_E}{2} + \log 2\pi\bigg] + \mathcal{O}(\epsilon) \nn
&&\mathcal{I}_{420}(\mu) = - \bigg(\frac{\Lambda}{4\pi \mu}\bigg)^{2\epsilon}\frac{1}{4\pi\mu^2}\bigg[\frac{5}{16} + \frac{5\big(-1 + \log 16 + 2\log\pi + \psi(0,7/2)\big)}{16}\epsilon \bigg] + \mathcal{O}(\epsilon^2)\nn
&&\mathcal{I}_{402}(\mu) = - \bigg(\frac{\Lambda}{4\pi \mu}\bigg)^{2\epsilon}\frac{1}{4\pi \mu^2}\bigg[\frac{1}{16} + \frac{-5 - 3\gamma_E + \log 64 + 6 \log \pi}{48}\epsilon \bigg] + \mathcal{O}(\epsilon^2)  \nn 
&&\mathcal{I}_{411}(\mu) = - \bigg(\frac{\Lambda}{4\pi \mu}\bigg)^{2\epsilon}\frac{1}{4\pi\mu^2} \bigg[ \frac{1}{16} + \frac{1 - 3\gamma_E + \log 64 + 6\log \pi}{48}\epsilon \bigg] + \mathcal{O}(\epsilon^2) ,
\label{SI-1}
\eea 
wherein the generalized polygamma function $\psi(z, q)$ appearing in the above sum integrals is given by 
$\psi(z,q) = e^{-\gamma z} \frac{\partial}{\partial z} \big[ e^{\gamma z} \frac{\zeta(z + 1, q)}{\Gamma(-z)}\big]$.

\end{document}